\documentclass[preprint2]{aastex}

\usepackage{lscape}
\usepackage{longtable,lscape}
\usepackage{natbib}
\usepackage{url}

\newcommand{\myeph}[1]{{{#1}}}
\newcommand{\mybf}{}
\newcommand{\mybfsec}{}

\shorttitle{A deep near-infrared survey toward Aquila $-$ I. Molecular hydrogen outflows}
\shortauthors{Zhang et al.}

\begin{document}


\title{A deep near-infrared survey toward the Aquila molecular cloud $-$ I. Molecular hydrogen outflows}


\author{Miaomiao Zhang\altaffilmark{1}, Min Fang\altaffilmark{1}, Hongchi Wang\altaffilmark{1}, Jia Sun\altaffilmark{1,2}, Min Wang\altaffilmark{1}, Zhibo Jiang\altaffilmark{1}, and Sumedh Anathipindika\altaffilmark{3}}
\affil{$^1$Purple Mountain Observatory, \& Key Laboratory for Radio Astronomy, Chinese Academy of Sciences, 210008 Nanjing, PR China}
\affil{$^2$University of Chinese Academy of Sciences, 100080 Beijing, PR China}
\affil{$^3$Indian Institute of Technology, Department of Physics, Kharagpur, India}







\begin{abstract}
We have performed an unbiased deep near-infrared survey toward the Aquila molecular cloud  with a sky coverage of $\sim$1 deg$^2$. \myeph{We identified 45 molecular hydrogen emission-line objects(MHOs), of which only 11 were previously known. Using the Spitzer archival data we also identified 802 young stellar objects (YSOs) in this region.} Based on the morphology and the location of MHOs and YSO candidates, we associate 43 MHOs with 40 YSO candidates. The distribution of jet length shows an exponential decrease in the number of outflows with increasing length and the molecular hydrogen outflows seem to be oriented randomly. Moreover, there is no obvious correlation between jet lengths, jet opening angles, or jet $H_2~1-0~S(1)$ luminosities and spectral indices of the possible driving sources in \myeph{this region}. \myeph{We also suggest that molecular hydrogen outflows in the Aquila molecular cloud are rather weak sources of turbulence, unlikely to generate the observed velocity dispersion in the region of survey.}
\end{abstract}





\section{Introduction}
Mass outflow plays an essential role in the process of star formation \citep{shu87,arce07,bally07}. It is believed that mass outflow is an important way to transfer the excess angular momentum from the dense molecular cores to the ambient interstellar medium \citep{shang07}. Outflows have been observed in different wavelengths: Herbig-Haro (HH) objects are the optical manifestation of shock-ionized mass outflows, which trace either material ejected from the protostars directly or the shocked interstellar medium \citep{rb01,wang04,hhperseus,wang05,wang06,wang09}. CO outflows detected in millimeter wavelength probe the swept-up or entrained medium {\mybf along the edges of the jets} \citep{bach96,wu04}.

In near-infrared bands, molecular hydrogen emission lines, in particular the $\nu=$1-0~S(1) transition at 2.12 $\mu$m, are powerful tracers of shock-excitation. \citet{davis10} has defined molecular hydrogen emission-line objects (MHOs) as the near-infrared manifestation of mass outflow and presented a comprehensive catalog of over 1400 MHOs. Wide-field deep near-infrared survey using a combination of narrow-band $H_2$ filter and the corresponding filter for continuum emission such as the $K_s$ filter has become a very efficient method to detect and identify MHOs \citep{stanke02,kha04,davis08,uwish2}. MHOs are also good tracers of young stellar objects (YSOs), in particular those deeply embedded in the molecular cloud cores. 
Once the relationship between the MHOs and their driving sources has been established, we can study the distribution of YSOs, the interaction of YSOs with the ambient interstellar medium, the star formation efficiency, and the evolution of the entire star-forming region by analyzing the statistical characteristics of the molecular hydrogen outflows \citep{stanke02,davis08,davis09,zhang13}. 

The Aquila Rift is located along the Galactic plane and the large Aquila Rift cloud complex stretches from 20\degr~to 40\degr~in longitude and -1\degr~to 10\degr~in latitude, as revealed by CO and H {\small I} observations \citep{dame01,aquilabook}. The molecular mass of the Aquila Rift estimated from CO observation is about 1.1-2.7$\times$10$^{5}$\,M$_{\sun}$ \citep{dame85,dame87,straizys03}. \citet{aquilabook} reviewed the Aquila Rift and compiled a list of known YSOs that includes only 9 YSOs from literature in the Aquila Rift. \myeph{The apparent lack of a large number of YSOs in Aquila is especially surprising given the age of young stars detected in this cloud \citep{prato03,rice06}, and the abundance of raw material for star-formation in this region. However, in the recent past, with the availability of data from the \textit{Spitzer}, WISE, and Herschel space telescopes, a number of new YSOs have been detected in this region.} 

The Aquila Rift {\mybf has now become a hot-spot} for the study of star formation. \citet{serpenssouth} discovered an embedded cluster of YSOs in the Serpens-Aquila Rift using the \textit{Spitzer} IRAC imaging data and identified 54 Class I and 37 Class II sources in the cluster. Similarly, as one of the targets of the Herschel Gould Belt key program \citep{andre10}, the Aquila region has also been surveyed with the PACS and SPIRE \citep{konyves10,bontemps10,men10}. \citet{bontemps10} used ``the Aquila Rift complex" to describe the molecular cloud that corresponds to a large extinction feature in the extinction map derived using the 2MASS catalog. The Aquila Rift complex mainly harbors two known sites of star formation \citep[see Fig. 1 in ][]{bontemps10}: Serpens South is the western young embedded cluster \citep{serpenssouth} and W40 is the eastern cluster associated with an H {\small II} region \citep{w40smith,vallee87}. ``The Aquila molecular cloud" \myeph{referred to in this paper is the same as the ``Aquila Rift complex" described by \cite {bontemps10}.} The distance estimates to the Aquila Rift complex vary from $\sim$200\,pc to up to 900\,pc \citep{radhak72,vallee87,straizys03,serpenssouth,w40book,shuping12}. \citet{bontemps10} compared the different estimates of distance and finally adopted the value of 260\,pc. In this paper, we follow their suggestion and also adopt the value of 260\,pc as the distance of the Aquila molecular cloud.

Hundreds of protostars have been discovered in the Aquila molecular cloud \citep{bontemps10}. \citet{nakamura11} identified 15 blueshifted and 10 redshifted outflow components using the CO (J $=$ 3-2) mapping observations in the Serpens South cloud. \citet{mho2213} identified several $H_2$ features around IRAS 18264-0143 in the Aquila region. \citet{tei12} identified 14 MHOs and associated them with 10 YSOs in the Serpens South cloud. 
\myeph{These works suggest that star-formation in the Aquila molecular cloud is still in its youth which makes this cloud an ideal laboratory to study various phases of stellar birth.}

In this paper we present our wide-field near-infrared survey toward the Aquila molecular cloud with a coverage of $\sim$1 square degree. Our goal is to detect the jets and outflows from  young stars in the Aquila molecular cloud. Compared with the optical observations, our near-infrared survey enables us to detect more outflow features due to the relatively lower extinction in near-infrared bands. \myeph{Combining our near-infrared observations of this region with the corresponding archival data from the \textit{Spitzer} and Herschel, we were able to associate H$_{2}$ emission features with their possible driving sources. We then studied the statistical characteristics of H$_{2}$ outflows in the Aquila molecular cloud. This paper is organized as follows :} In Section~\ref{obsdata} we describe our near-infrared observations and the complementary archival data. We then present our results in Section~\ref{results}, including the \myeph{scheme employed to identify outflows and YSOs.} In Section~\ref{discuss}, we discuss the characteristics of molecular hydrogen outflows in the Aquila molecular cloud, including a statistical analysis of outflow parameters such as jet lengths, jet opening angles, jet orientations and the H$_{2}$ (1-0) S(1) luminosity of jets. We also \myeph{calculate the momentum injected by molecular hydrogen outflows in this cloud and critically evaluate its contribution to turbulence that appears to support the Aquila molecular cloud against self-gravity.} Finally we summarize our findings in Section~\ref{summary}. 

\section{Observations and data reduction}\label{obsdata}
\subsection{Near-infrared imaging}
The observations were conducted in queue-scheduled observing (QSO) mode between 26th and 29th July, 2012 with WIRCam \citep{wircam} equipped on {\mybf Canada-France-Hawaii Telescope (CFHT)}, covering in total an area of $\sim$1 deg$^2$. WIRCam is a near-infrared mosaic imager with four 2048$\times$2048 CCDs, yielding a field of view of $\sim$21\arcmin$\times$21\arcmin~with a plate scale of 0\farcs306 pixel$^{-1}$. We observed 10 fields toward the Aquila molecular cloud in $J$, $H$, $K_{s}$, and $H_{2}$ bands. 
The narrowband $H_{2}$ filter has a central wavelength of 2.122\,\micron~and a bandwidth of 0.032\,\micron\footnote{The information of the broadband filters can be found at \url{http://www.cfht.hawaii.edu/Instruments/Filters/wircam.html}}. Each field was imaged with a four point dithering pattern. The number of sub-exposures per dithering pattern position for $J$, $H$, $K_s$, and $H_2$ band is 2, 3, 2, and 3, respectively, while the individual exposure time is 30\,s in $J$ band, 10.3\,s in $H$ band, 15\,s in $K_s$ band, and 100\,s in $H_2$ band. Thus the total integration time of $J$, $H$, $K_s$, and $H_2$ band for each field is 240s, 123.6s, 120s, and 1200s, respectively. Figure~\ref{covershow} shows the coverage of our CFHT/WIRCam observations with red boxes. 

Individual images are primarily processed by the CFHT `I'iwi pipeline (the IDL Interpretor of the WIRCam Images, version 2.1.1), which includes dark subtraction, flat-fielding, non-linearity correction, cross-talk removal,  and sky subtraction\footnote{The details of the pipeline can be found at \url{http://www.cfht.hawaii.edu/Instruments/Imaging/WIRCam/IiwiVersion1Doc.html}}. Then, the data are handled with the SIMPLE IDL package\footnote{\url{http://www3.asiaa.sinica.edu.tw/~whwang/idl/SIMPLE/index.htm}}, which is an IDL based data reduction package for optical and near-IR blank-field imaging observations. We use SIMPLE-WIRCAM (SIMPLE Imaging and Mosaicking PipeLinE for WIRCAM) to remove the distortion and do the absolute astrometry on a single frame by comparing the image with the 2MASS reference catalog \citep{2mass}. To estimate the accuracy of our astrometry, we use Sextractor \citep{sex} to do source detection in each filter of each field and compare the coordinates of detected sources with the WCS entries of 2MASS catalog. We find that the root-mean-square (rms) values of the difference of coordinates between the sources on the final mosaic images and 2MASS catalog for all fields are below 0.3\arcsec. Based on the accurate astrometry, the dithered individual exposures are finally combined into the stacked images with the software SWARP \citep{swarp}, which is a program that resamples and co-adds together FITS images using any arbitrary astrometric projection defined in the WCS standard\footnote{\url{http://www.astromatic.net/software/swarp}}. 

Point source detection and aperture photometry\footnote{As our survey region is close to the Galactic plane, the more reliable PSF fitting method for source detection and photometry is still under construction. By comparing the results from aperture photometry and PSF fitting for one field that is closest to the Galactic plane, we found that the mean value and root mean square value of difference of photometry between these two methods are $<$0.02 mag and $<$0.07 mag individually for all bands, but that PSF fitting method can detect larger number of faint sources. In this paper, photometry for the point sources in our near-infrared images is only used to select YSO candidates as the supplement of \textit{Spitzer} data in the YSO selection scheme suggested by \citet{gutermuth09}. Thus the aperture photometry is enough to match our science objective. The PSF fitting results and follow-up analysis will be published in our subsequent paper (Sun et al. 2015, in preparation).} is performed on the mosaic images via the IDL routine {\it find} and {\it aper} with an aperture radius of 1.5\arcsec~and concentric sky annuli of inner and outer radii of 3.0\arcsec~and 4.5\arcsec~respectively. Photometric zero points and color terms were calculated by comparison of the instrumental magnitudes of relatively isolated, unsaturated bright sources with the counterparts in the 2MASS Point Source Catalog \citep{2mass}. By comparing $\sim$23000 sources detected in both the CFHT and the 2MASS observations, we find that the photometric reliability in all bands is $\sim$0.08-0.16 mag, depending on the source brightness.


\subsection{Spitzer data}
\begin{figure*}
\centering
\includegraphics[width=12cm]{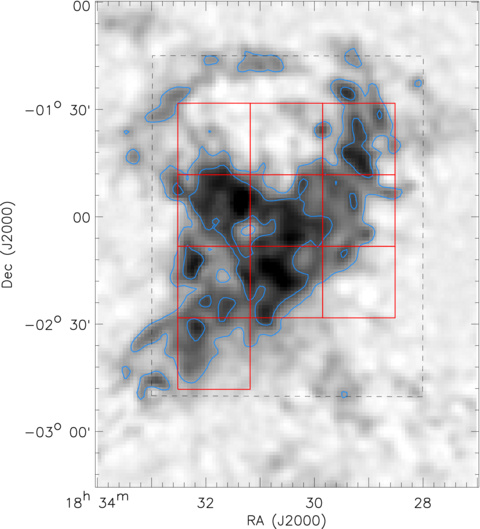}
\caption{Coverage of our data. The underlying image is the extinction map from \citet{dobashi11}. The blue contours represent the visual extinction of 5 and 10 magnitudes. Ten fields observed with CFHT/WIRCam are marked with red boxes while the black dashed box shows the coverage of our retrieved {\it Spitzer} archive data.}
\label{covershow}
\end{figure*}

As part of the {\it Gould Belt Legacy program} (PID: 30574), the {\it Spitzer} Space Telescope observations toward the Serpens-Aquila rift were conducted in May and September, 2007 with the IRAC and MIPS cameras \citep{irac,mips}. IRAC images at 3.6, 4.5, 5.8, and 8.0\,\micron~were made in High Dynamic Range mode with integration times of 0.4 and 10.4\,s. We download the IRAC standard basic calibrated data (BCD) products provided by the {\it Spitzer} Science Center from their standard data processing pipeline version S18.18. Final mosaics are built at the native instrument resolution of 1.2\arcsec~pixel$^{-1}$ for each Astronomical Observation Request (AOR), using {\it Mopex}\footnote{\url{http://irsa.ipac.caltech.edu/data/SPITZER/docs/dataanalysistools/tools/mopex/}} \citep[version 18.5, ][]{mopex}, which is a package for reducing and analyzing imaging data. Thus for each AOR, we obtain two IRAC mosaics, one built from the long exposures of 10.4\,s and the other built from the short exposures of 0.4\,s. MIPS images at 24, 70, and 160\,\micron~were obtained at the fast scan rate of 17\arcsec~s$^{-1}$. We also download the MIPS post-BCD data products provided by the {\it Spitzer} Science Center from their pipeline version S18.12. Note that we only use the MIPS 24\,\micron~images with the resolution of 2.45\arcsec~pixel$^{-1}$ in this paper. The {\it Spitzer} observations toward the Serpens-Aquila rift cover a very large sky region. To match with our near-infrared imaging data, we finally restrict the IRAC and MIPS data to the region that is marked with the black dashed box in Fig.~\ref{covershow}. Figure~\ref{spitzercover} (left panel) shows the three-color image that is constructed with IRAC 3.6\,\micron~(blue), 4.5\,\micron~(green), and 8.0\,\micron~(red) images.

Point source detection and aperture photometry is performed on the final IRAC and MIPS mosaics via the photometry and visualization tool, PhotVis \citep[version 1.10; ][]{gutermuth04,gutermuth08}. Aperture photometry is performed on the IRAC images with an aperture radius of 2.4\arcsec~and background flux is estimated with an concentric sky annuli of inner and outer radii of 2.4\arcsec, and 7.2\arcsec~respectively. Aperture and inner and outer sky annulus radii are selected to be 7.6\arcsec, 7.6\arcsec, and 17.8\arcsec, respectively, for MIPS 24\,\micron~images. Photometric zero points for IRAC aperture photometry are derived from the calibrations presented in \citet{reach05}, including standard aperture corrections for the radii adopted. The photometric zero point of MIPS 24\,\micron~band is the suggested value from \citet{gutermuth08}. The 90\% completeness limits are derived by adding successively dimmer sets of {\it Spitzer} PRFs (Point Response Functions\footnote{see \url{http://irsa.ipac.caltech.edu/data/SPITZER/docs/irac/calibrationfiles/psfprf/} and \url{http://irsa.ipac.caltech.edu/data/SPITZER/docs/mips/calibrationfiles/prfs/}}), extracting their fluxes using the same procedure described above. The detection completeness limits are defined as the magnitudes at which 90\% of the synthetic stars are recovered. Our 90\% completeness limits are 14.7, 15.1, 13.4, 13.0, and 8.2 mag at 3.6, 4.5, 5.8, 8.0, and 24\,\micron. 

\subsection{Multi-band photometric catalog}\label{bandmerge}
To obtain the final multi-band photometric catalog, we performed the bandmerge in four stages. First, we crossmatch our CFHT photometric catalog with the 2MASS point source catalog \citep{2mass} using a tolerance of 2\arcsec. The photometry of bright sources from our CFHT photometric catalog is contaminated by saturation. Thus for the point sources brighter than $J=14.5$, or $H=13.5$, or $K_s=13$ mag we use 2MASS photometric results. Second, we combine the long exposure photometry and short exposure photometry for each IRAC band and then the four IRAC band source lists are matched with a tolerance of 2\arcsec. Third, the MIPS 24\,\micron~catalog is integrated with a wider tolerance of 4\arcsec~due to the low resolution (6\arcsec) of the MIPS 24\,\micron~band. Finally, we crossmatch the {\it Spitzer} photometric catalog and the near-infrared photometric catalog with a tolerance of 2\arcsec. Note that the final catalog uses the mean IRAC and MIPS1 positions for all entries. 

\subsection{Herschel extinction map}\label{extmap}
The Herschel archival data used in this paper are part of the Herschel Gould Belt guaranteed time key programmes for the study of star formation with the PACS \citep{pacs} and SPIRE \citep{spire} instruments \citep{andre10} and have been published in \citet{andre10}, \citet{konyves10}, and \citet{bontemps10}.  The details of the observations and data reduction can be found in \citet{konyves10}. We use the final calibrated images\footnote{\url{http://www.herschel.fr/cea/gouldbelt/en/Phocea/Vie_des_labos/Ast/ast_visu.php?id_ast=66}} at 160, 250, 350, and 500\,\micron~with the angular resolutions of $\sim$12\arcsec, $\sim$18\arcsec, $\sim$25\arcsec, and $\sim$37\arcsec, respectively. 

All Herschel maps are smoothed to the beam size of the 500\,\micron~map, through convolving the maps with the convolution kernels\footnote{\url{http://dirty.as.arizona.edu/~kgordon/mips/conv_psfs/conv_psfs.html}} supplied by \citet{gordon08}. The column density was determined from a pixel-to-pixel modified black body fit to four longer wavebands of PACS and SPIRE from 160\,\micron~to 500\,\micron, assuming the dust opacity law of \citet{beckwith90} and dust emissivity index of 2 \citep{hilde83}, following the same procedure described by \citet{konyves10}. Note that we do not use the PACS 70\,\micron~map to construct the column density map because 70\,\micron~data may not be entirely tracing the cold dust, as suggested by \citet{hill11}. The column density map is calibrated with the 2MASS extinction map from \citet{dobashi11}. The final Herschel extinction map is obtained using the relation $N_{H_2} = 10^{21}\times A_V$ \citep{bohlin78,konyves10}. Figure~\ref{spitzercover} (right panel) shows the obtained Herschel extinction map of the Aquila molecular cloud, overlaying the contours with levels of $A_V=4$, 10, and 30 mag.


\section{Results}\label{results}
\subsection{The detected MHOs in Aquila}
For each field, we have obtained the $H_2$ narrow-band image and the continuum $K_s$ band image. We use Sextractor \citep{sex} to detect point sources on $H_2$ images and $K_s$ images. Then the unsaturated bright point sources are used to adjust the flux level of stars in the $H_2$ image and $K_s$ image and a continuum-subtracted $H_2$ image ($H_2-K_s$) is obtained for each field.

We did the visual inspection on all $H_2-K_s$ images to search for {\mybf $H_2$ line emission features.}
 These {\mybf $H_2$ line emission features} are treated as MHO candidates. In order to avoid the inclusion of instrumental artifacts, each MHO candidate is examined in the corresponding $H_2$, $J$, $H$, and $K_s$ images. Any potential artifact has been removed from the MHO candidate list. 

Emission nebulae, such as planetary nebulae (PNe), supernova remnants (SNRs) and H {\small II} regions may contaminate our identification. PNe usually exhibits symmetrical morphology and can be distinguished from MHOs based on their morphological difference. {\mybf We found no PNe in our MHO candidate list.} Moreover, the known H {\small II} region, W40, is located in our survey region of the Aquila molecular cloud. We examined MHO candidates near W40 and {\mybf two ($\sim$2\%)} potential contaminants have been removed from our MHO candidate list. 
The remaining extended $H_2$ emission features are identified as MHOs.

%


{\mybf Finally, we have identified 45 MHOs that consist of 108 discrete $H_2$ emission-line features. All are situated within the Aquila molecular cloud (see Fig.~\ref{outflows}). Of these 45 MHOs, 11 are previously known objects while 34 are newly discovered. Table~\ref{mhofeatures} in the appendix lists the position, surface area, radius, and integrated $H_2$ line flux for each sub-feature within each MHO. The MHO numbers assigned to each object have been entered into the MHO catalog\footnote{{\mybfsec \url{ http://astro.kent.ac.uk/~df/MHCat/}}} \citep{davis10}. A description of each MHO is also given in the appendix.}

\subsection{Photometry of MHO features}\label{photmho}
Areal photometry is performed to measure the fluxes of MHO features. We define manually a polygonal aperture based on the morphology and surface brightness distribution of each MHO feature. The principle for the aperture definition is to ensure that no stars are in the aperture and the aperture {\mybf contains as little 
background area outside the bounds of each $H_2$ emission feature as possible}. Note that there are circular holes in some polygonal apertures in order to avoid stars. Figs.~\ref{figa01}-\ref{figa42} in the appendix show these apertures with blue polygons and the holes are also marked with red circles. For each MHO feature, a same polygon aperture toward the nearby sky region that is emission free is used to estimate the local sky background. The polygon apertures around the $H_2$ features and the nearby sky apertures are applied to the MHO features on the continuum-subtracted $H_2$ images to measure the fluxes of the $H_2$ 1-0 S(1) emission line.

The flux calibration is obtained via our near-infrared photometric catalog (2MASS$+$CFHT, see section~\ref{bandmerge}). We use Sextractor \citep{sex} to measure the fluxes of the point sources on the $H_2$ images. The zero-point flux is derived through the comparison of the integrated counts with the cataloged flux of the point sources. We then applied the zero-point flux to the integrated counts of the MHO features with the aim to convert the counts of MHO features to the flux unit. Uncertainties in the areal photometry are estimated from the variation of the local sky background level in the continuum-subtracted $H_2$ images.

The obtained fluxes and uncertainties of the MHO features are shown in table~\ref{mhofeatures} in the appendix. In our $H_2$ images, outflow features with a surface brightness of $\sim$1.3$\times$10$^{-19}$ W m$^{-2}$ arcsec$^{-2}$ are detected at $\sim$5$\sigma$ above the surrounding background. The median of the fluxes of MHO features is 1.2 $\times$ 10$^{-17}$ W\,m$^{-2}$ and the minimum of the fluxes we detected is 6.6 $\times$ 10$^{-19}$ W\,m$^{-2}$. 

\subsection{Identification of YSOs in Aquila}
{\mybf YSOs usually show excessive infrared emission that can be used to \myeph{distinguish them} from field stars and classify \myeph{them into} different evolutionary classes. \citet{lada84} and \citet{lada87} developed an empirical classification scheme that firstly codified a tripartite class system based on the slope of the SED to indicate the different evolutionary stages of YSOs, which has become a `standard' classification system of YSOs. Some authors use the infrared spectral index to identify YSO candidates in the star-forming regions \citep{mallick13,kim15}. YSOs \myeph{can also be classified using their mid-IR clours} (i.e., color-magnitude diagrams). 
\citet{gutermuth08,gutermuth09} established a mid-IR color-based method to identify YSOs, which has been widely applied to the YSO identification in the star-forming regions. \myeph{We use this scheme in the present work as it can efficiently mitigate the effects of contamination from field stars and extragalactic sources. We calculate the slope of SEDs and estimate the evolutionary stages of the YSOs detected in this survey using the classification scheme suggested by \citet{lada87}.}}

YSO identification is based on our multi-band photometric catalog (see section~\ref{bandmerge}) and the source classification scheme presented in \citet{gutermuth08,gutermuth09}. The details about this multiphase source classification scheme can be found in the appendix of \citet{gutermuth09}. {\mybf Here we just summarize our process (also see the description in \citet{rapson14}).

There are three phases in the YSO selection scheme suggested by \citet{gutermuth09}.
\textit{Phase 1} is applied to the sources that have detections in all four IRAC bands with photometric uncertainties $\sigma<$0.2 mag. After removing the contaminants such as star-forming galaxies, broad-line AGNs, unresolved knots of shock emission, and sources that have PAH-contaminated apertures, the sources with [4.5]-[5.8] $>$ 0.7 mag and [3.6]-[4.5] $>$ 0.7 mag are classified as Class I candidates while the sources with [4.5]-[8.0] $>$ 0.5 mag, [3.6]-[5.8] $>$ 0.35 mag, [3.6]-[4.5] $>$ 0.15 mag, and [3.6]-[5.8] $\leq$ 3.5$\times$([4.5]-[8.0]-0.5)$+$0.5 mag are considered as Class II candidates, accounting for photometric uncertainty. The remaining sources are classified as Class III/field sources.

\textit{Phase 2} is applied to the sources that lack detections at either 5.8 or 8.0 \micron, but have high quality ($\sigma<$0.1 mag) near-infrared detections in J, H, and Ks bands. In order to distinguish the sources with IR-excess from those that are simply reddened by dust along the line of sight, we deredden the photometry of sources based on the extinction law presented in \citet{jiang14} and \citet{extinction07}. The dereddened $K_s-$[3.6] and [3.6]-[4.5] colors are used to identify the sources with infrared excess at 3.6\,\micron~and 4.5\,\micron~accounting for photometric uncertainty. Sources with no IR excess at 3.6\,\micron~and 4.5\,\micron~are presumed as Class III/field sources.

\textit{Phase 3} is applied to the sources that have detections in MIPS 24\,\micron~band with the photometric uncertainties $\sigma<$0.2 mag. The Class III/field sources that were classified in previous two phases are re-examined and sources with colors of [5.8]-[24] $>$ 2.5 mag or [4.5]-[24] $>$ 2.5 mag and [3.6] $<$ 14 are classified as transition disk candidates. For the sources that lack detection in some IRAC bands, but are very bright at 24\,\micron, i.e., [24] $<$ 7 mag and [IRACX]-[24] $>$ 4.5 mag, they are classified as deeply embedded Class I protostars, where [IRACX] is the photometric magnitude for the longest wavelength IRAC detection. The AGN candidates and shock emission dominated sources that were classified as contaminants in phase 1 are also re-examined and identified as Class I candidates if they have both bright MIPS 24\,\micron~photometry ([24] $<$ 7 mag) and convincingly red IRAC/MIPS colors of [3.6]-[5.8] $>$ 0.5 mag and [4.5]-[24] $>$ 4.5 mag and [8.0]-[24] $>$ 4 mag. Finally, the Class I source that were classified in all three phases
are re-analyzed to ensure their nature. Sources with [5.8]-[24] $>$ 4 mag or [4.5]-[24] $>$ 4 mag are classified as true Class I sources while other sources that do not match the above requirements are identified as highly reddened Class II sources. Figure~\ref{ccdyso} (left panel) shows the color-color diagrams that display the results of applying the above classification criteria to our multi-band photometric catalog.}

Finally we have identified 802 YSO candidates {\mybf using the above criteria \citep{gutermuth09}}, including {\mybf 151} Class I sources, {\mybf 629} Class II sources, and 22 transition disks. Figure~\ref{spitzercover} (left panel) shows the spatial distribution of these YSO candidates. 

The source spectral index, the slope of the source's spectral energy distribution (SED) that is defined as
\begin{displaymath}
\alpha = \frac{dlog(\lambda S(\lambda))}{dlog(\lambda)}
\end{displaymath}
where $S(\lambda)$ is the source's flux density at wavelength $\lambda$, can be used to determine the evolutionary state of a source. We calculate the spectral indices of all 802 YSOs through fitting their observed SEDs from 2\,\micron~to 24\,\micron. Using the YSO classification scheme suggested by \citet{lada87}, we reclassify these 802 YSOs into {\mybf 193} Class I sources, {\mybf 601} Class II sources, and 8 Class III/MS sources.

The Aquila molecular cloud is highly obscured. The mean visual extinction in this region estimated from our Herschel extinction map is $\sim$5 magnitude. Thus for the YSO candidates obtained above, it is desirable to correct their flux densities for extinction. However, due to the lack of spectral types of YSO candidates, we can only roughly determine their extinction.

For the YSO candidates with detections in $J$, $H$, and $K_s$ bands, we adopt the method suggested by \citet{fang13}, in which the extinction is obtained by employing the $H - K_s$ versus $J - H$ color-color diagram. The detailed description of this method can be found in \citet{fang13}. Here we just summarize some \myeph{only a few aspects of this scheme}. The location of each YSO in the $H - K_s$ versus $J - H$ color-color diagram depends on both its intrinsic colors and its extinction. Figure~\ref{ccdyso} shows the $H - K_s$ versus $J - H$ color-color diagram of the YSO candidates in the Aquila molecular cloud. Given the different origins of intrinsic colors of YSOs, we divide the color-color diagram into three sub-regions. Different methods are used to obtain the intrinsic color of the YSOs in different regions: in region 1, 
the intrinsic color $[J - H]_{0}$ values of YSOs are simply assumed to be 0.6, which is the typical value of a K5-type dwarf star; in region 2, the $[J - H]_{0}$ value of a YSO candidate is obtained from the intersection between the reddening vector and the locus of main-sequence stars from \citet{bb88}; in region 3, the intrinsic $[J - H]$ color is derived from where the reddening vector and the CTTS locus \citep{meyer97} intersects. Then the extinction values of individual YSOs are estimated from their observed color $[J - H]_{obs}$ and their intrinsic color $[J - H]_{0}$, using the equation $A_K = ([J - H]_{obs} - [J - H]_{0})/(A_J/A_K-A_H/A_K)$, where the {\mybf extinction ratios} \myeph{were adopted} from \citet{jiang14}.

For the YSO candidates outside these three regions or without detection in $J$, $H$, or $K_s$ bands, their extinction is estimated from our Herschel extinction map. We firstly obtain the mean values of nine beam-size ($\sim$37\arcsec~per beam) pixels around the YSO candidates and then use the half of these mean values as the extinction values of individual YSO candidates, assuming that all the YSO candidates are located in the middle of the Aquila molecular cloud along the line of sight. {\mybfsec Here we must stress that it is very difficult to accurately derive the extinction towards YSOs even if their spectral types are known because of the contribution from the disk and accretion flow. Thus the specific value of extinction towards some YSO candidates obtained using the above methods could have a large uncertainty.}

\myeph{For colour excesses in the near-infrared we adopted the extinction law suggested by \citet{jiang14}, while adopting the one suggested by \citet{extinction07} for colour excesses in the mid-infrared.} The median value of visual extinction toward the YSO candidates is 11 magnitudes. Based on the de-reddened flux densities of 802 YSO candidates, we calculate their de-reddened spectral indices and reclassify them according to their de-reddened spectral-index, $\alpha$. Finally, we obtain {\mybf 100} Class I sources, {\mybf 551} Class II sources, and {\mybf 151} Class III sources. Figure~\ref{yso_alpha} shows the distributions of apparent and de-reddened spectral indices of 802 YSO candidates.


\subsection{Driving sources of MHOs in Aquila}
The likely driving sources of MHOs were identified \myeph{on the basis of the morphologies and locations of individual MHOs relative to the candidate YSOs using which, we associated 43 MHOs with 40 YSO candidates that led us to identify 40 molecular hydrogen outflows.} Table~\ref{outflows_info} lists the information of these 40 molecular hydrogen outflows. 
Figure~\ref{outflows} shows the spatial distribution of the molecular hydrogen outflows with blue lines that connect the MHOs and their likely driving sources. Note that there are two MHOs (MHO 3284-3285) whose driving sources can not be identified due to the lack of YSOs around them. The details about the driving source identification can be found in the appendix.


Of 40 outflow sources, 30 (75\%) are protostars ($\alpha>$0) when we use the apparent spectral indices to classify the YSOs. After correcting fluxes of outflow sources for extinction, 27 ($\sim$68\%) of 40 outflow sources belong to protostars. Thus $\sim$70\% $H_2$ outflows in the Aquila molecular cloud are driven by protostars. This fraction is lower than the value of $>$80\% in Orion A \citep{davis09} and higher than the value of $<$50\% in Ophiuchus \citep{zhang13}. The mean value of the apparent spectral indices of outflow sources for Aquila is $\sim$0.6, which is lower than the value of 0.86 in Orion A \citep{davis09} and higher than the value of -0.16 in Ophiuchus \citep{zhang13}. We note that the detection limit for $H_2$ features ($\sim$1.3$\times$10$^{-19}$ W m$^{-2}$ arcsec$^{-2}$) of our survey toward the Aquila molecular cloud is better than that toward Orion A \citep[7$\times$10$^{-19}$ W m$^{-2}$ arcsec$^{-2}$,][]{davis09} and similar to that toward Ophiuchus \citep[1$\times$10$^{-19}$ W m$^{-2}$ arcsec$^{-2}$,][]{zhang13}. The difference in the mean value of the apparent spectral indices \myeph{is likely due to various selection effects discussed by \citet{zhang13}.} With better sensitivity, we detect more low power jets from Class II sources in the Aquila molecular cloud than in Orion A. As the Aquila molecular cloud is located at a distance of $\sim$260 pc, farther than Ophiuchus \citep[120 pc,][]{lombardi08,loinard08}, less outflows with small spatial extent will be detected in Aquila than in Ophiuchus when the surface brightness detection limits are similar. 

We can also estimate that about {\mybf 16\%}~of protostars in our survey region drive $H_2$ outflows before de-reddening. After de-reddening for YSOs, the fraction of protostars that drive $H_2$ outflows turns out to be {\mybf 27\%}; \myeph{this fraction is respectively, 30\% for the Orion A \citep{davis09}, and~20\% for the Ophiuchus molecular clouds \citep{zhang13}.}  We also find that, of our 40 outflow sources, 18 are associated with infrared nebulae. Figure~\ref{nebshow} shows the $K_s-$band images of these 18 infrared nebulae 
Note that several outflow sources are invisible in near-infrared.

\subsection{Interesting individual objects}
\subsubsection{A disk-jet system}
Figure~\ref{jet-disk} shows the region \myeph{in the vicinity of the} MHO 3251b1 and its driving source ID\#17, which constitute a disk-jet system. The left panel is the $H_2$ image and the right panel is the $K_s$ image. The solid line in the left panel shows the collimated jet, whose bright head has been identified as MHO 3251b1 (see Fig.~\ref{figa07} in Appendix). It seems that there are several faint knots to the northwest of MHO 3251b1 along the solid line, but these structures are too faint \myeph{to be identified with any degree of certainty.} Another obvious structure in Fig.~\ref{jet-disk} is the bipolar infrared nebula. A dark lane \myeph{is also visible between} the two bright lobes and the orientation of the dark lane is roughly perpendicular to the jet. The white plus in the right panel marks the position of ID\#17, which is the mean position of IRAC and MIPS 24\,\micron~weighted with flux in individual band. We also mark the mean position of IRAC for ID\#17 using the blue cross and the MIPS 24\,\micron~position using the red cross. We note that all positions of ID\#17 are roughly located in the dark lane, indicating a protostar surrounded by an edge-on disk. The apparent size of the dark lane at the base of the northern lobe of the nebula is $\sim$4\arcsec, corresponding to $\sim$1000 AU for an assumed distance of 260 pc. The physical size of the bipolar nebula is $\sim$1800 $\times$ 1300 AU, which is $\sim$2 times larger than those for the detected edge-on systems in Taurus \citep{pad99}.

\subsubsection{An H$_2$ outflow from a Class 0 source}
Figure~\ref{jet-protostar} shows the region of MHO 3260 (also see Fig.~\ref{figa12}) which is an $H_2$ outflow from ID\#2 that is identified as a Class 0 source by \citet{maury11}. The top-left panel is a three-color image constructed with CFHT $J$ (blue), $H_2$ (green), and $K_s$ (red) images and MHO 3260 exhibits as green features in the panel. The detailed description about MHO 3260 can be found in the appendix. The top-right panel shows the three-color image constructed with IRAC 3.6\,\micron~(blue), 4.5\,\micron~(green), and 8.0\,\micron~(red) images. {\mybf The IRAC 4.5\,\micron~ band contains $H_2$ ($\nu$ = 0 $-$ 0, S(9, 10, 11)) lines and CO ($\nu$ = 1 $-$ 0) band heads \citep{reach06} and has proved an effective tool for exhibiting the mid-infrared emission from outflows.} Some green features that correspond to MHO 3260 can be seen in this panel. These extended green objects (EGOs) represent the shock-excited emission at 4.5\,micron, which are usually suggested as mid-infrared outflows \citep{ego08,cham09,zw09}. The bottom-left panel is the three-color image constructed with MIPS 24\,\micron~(blue), PACS 70\,\micron~(green), and 160\,\micron~(red) images while the bottom-right panel is the three-color image constructed with SPIRE 250\,\micron~(blue), 350\,\micron~(green), and 500\,\micron~(red) images. We note that ID\#2 is invisible at 1.2-24\,\micron~and visible in Herschel 70-500\,\micron~bands. Actually, \citet{maury11} also detected the counterpart of ID\#2 in 1.2 mm dust continuum image and classified it as a Class 0 source. Thus MHO 3260 is an $H_2$ outflow driven by a Class 0 source.

\section{Discussion}\label{discuss}
\subsection{Outflow statistics}
Based on the positions and photometry of MHO features, we calculate important outflow parameters such as jet length ($L$), jet opening angle ($\theta$), jet position angle (PA), and jet $H_2~1-0~S(1)$ line luminosity according to the suggestions from \citet{davis09,ioa1,ioa2,zhang13}. 

A jet opening angle ($\theta$) is measured from a cone with the smallest vertex angle centered on the driving source that {\mybf includes all of the detected $H_2$ line emission associated with each MHO outflow in our continuum-subtracted $H_2$ image.} Thus the opening angle can be obtained even for the outflow that consists of only one feature. A jet position angle (PA) is measured {\mybfsec east of north} as computed from the bisector of the opening angle. For the bipolar outflows, we select the angles smaller than 180 degrees as the position angles. A jet length ($L$) is defined as the maximal distance from the pixels of MHO features to the source projected to the bisector of the opening angle. For the bipolar outflows, we sum up both lengths over two lobes. We have obtained the flux for each MHO feature in section~\ref{photmho}. The flux of each molecular hydrogen outflow is obtained through summing up all fluxes over the MHO features in the outflow. Then the $H_2~1-0~S(1)$ luminosity of each outflow is calculated using the relation of $L_{H_2} = F_{H_2} \times 4\pi d^2$, assuming the distance of 260\,pc for the Aquila molecular cloud. Note that we did not consider the effect of extinction. Table~\ref{outflows_info} lists these parameters and Fig.~\ref{parahist} shows the distribution of these outflow parameters. Note that these parameters are only calculated with MHOs (not considering the information provided by HH objects and CO outflows) and uncorrected for inclination to the line of sight.

\subsubsection{Jet lengths}
We have obtained the projected lengths for all 40 molecular hydrogen outflows in the Aquila molecular cloud (see table~\ref{outflows_info}). The distribution of jet lengths shows a steep decrease in the number of outflows with increasing length in Fig.~\ref{parahist}. In our sample, there are two outflows (5\%) longer than 1\,pc, 30 outflows (75\%) longer than 0.1\,pc. It seems that the fraction of parsec scale outflows in our sample (5\%) is similar to that of $\sim$8-9\% in Orion A \citep{stanke02,davis09}, but smaller than $\sim$12\% in the west of Perseus \citep{davis08}. The median length of our sample is 0.24\,pc, with the maximum of 1.65\,pc. 

\myeph{Although the distribution of jet lengths in our sample can be fitted with both, an exponential and a power-law, we find that the distribution is closer to an exponential behavior than a power-law function. Shown in Fig.~\ref{jetlength} is a histogram with logarithmic axes of the jet lengths; the solid line here represents a linear fit to this distribution.} The dotted lines show the linear fittings for the distributions of jet lengths in different bin sizes. The slopes obtained from the different fittings are from -0.66 to -1.03, depending on the bin size. Here we adopt a mean value of slopes to describe the behavior of number N of outflows with flow length in the following way:
\begin{displaymath}
N \propto 10^{(-0.85\pm0.1)\times length(pc)}
\end{displaymath}
where the uncertainty of slope is the standard deviation of all fitting slopes. This value of the exponent, $-0.85$, is \myeph{consistent with the value obtained by \citet{ioa2}.} \citet{ioa1,ioa2} detected 134 MHOs in the Galactic plane toward Serpens and Aquila using the UKIRT telescope and identified 131 molecular hydrogen outflows. They fitted the distribution of jet lengths with the exponential function and obtained a slope of $-0.75$. The outflows detected by \citet{ioa1,ioa2} are not associated with the Aquila molecular cloud due to the large distances of $>$2\,kpc of outflows in their sample. Their survey traces more luminous outflows and about 25\% of outflows in their sample are parsec-scale flows. Thus it seems that outflows from low-mass stars and massive/intermediate stars may have the similar behavior in projected lengths.

Figure~\ref{paravsalpha} (left two panels) shows the relation between jet lengths and the spectral indices of driving sources. Note that outflow source ID\#2 is invisible at 2-24\,$\mu$m while the outflow source ID\#19 is saturated in all IRAC bands and MIPS 24\,\micron~band. To plot these sources on the figure, we assume the value of 3 for their spectral indices because they are associated with the protostars identified by \citet{bontemps10}. Two longest outflows are both driven by Class I sources. Actually, the mean length of outflows from Class I sources is 0.34\,pc while the mean length of outflows from Class II/III sources is 0.22\,pc if using the apparent spectral indices to do the YSO classification. After we correct the spectral indices for extinction, these two values become 0.36\,pc and 0.22\,pc. It seems that Class I sources drive longer flows than Class II/III sources. However, from Fig.~\ref{paravsalpha} we can not find any obvious correlation between jet lengths and spectral indices of driving sources (correlation coefficients $<$ 0.3). Of course, we can not expect a simple linear correlation between jet lengths and spectral indices of driving sources because the relation between them should be complex: for the very young sources, we should expect a short jet length as a jet should start from the length of zero; then jets probably expand, and later on get shorter again as force support fades \citep{phdstanke,zhang13}. However, we can neither see any non-linear relation between jet lengths and spectral indices of driving sources from Fig.~\ref{paravsalpha}. 
Similar results have been found in other star-forming regions such as Orion A \citep{davis09}, Perseus \citep{davis08}, and Ophiuchus \citep{zhang13}. The molecular hydrogen outflows in Orion A, Perseus, and Ophiuchus also show no correlation between jet lengths and outflow source spectral indices.

\subsubsection{Jet opening angles}
Figure~\ref{parahist} (top right) shows the distribution of jet opening angles of outflows detected in the Aquila molecular cloud. The median opening angle of our outflow sample is 31\degr, with a minimum of 1.3\degr~and a maximum of 149\degr. 

Jet opening angle can be used to describe the collimation of an outflow. Some millimeter observations indicate that protostars that are in an earlier stage of evolution have \myeph{a higher likelihood of driving more collimated} CO outflows \citep{lee02,as06}. However, this trend has not been found in the molecular hydrogen outflow samples. \citet{davis08,davis09} and \citet{zhang13} investigated the relation between $H_2$ jet opening angles and spectral indices of driving sources in Perseus, Orion A, and Ophiuchus, respectively, and they all found no obvious correlation between jet opening angles and the spectral indices of driving sources. Figure~\ref{paravsalpha} (middle two panels) shows the relation between jet opening angles and spectral indices of driving sources in our outflow sample. The correlation coefficients between jet opening angles and apparent or de-reddened spectral indices of driving sources are both $<$0.3. However, the mean opening angle of outflows driven by Class I sources in our sample is 48\degr~while the mean opening angle of outflow driven by Class II/III sources is 22\degr~if using apparent spectral indices as the YSO classification criteria. If we use the de-reddened spectral indices to do the YSO classification, the mean opening angles of outflows driven by Class I sources and Class II/III sources become 51\degr~and 20\degr, respectively. {\mybf Despite the lack of a linear correlation} between jet opening angles and outflow source spectral indices, it seems that Class I sources in the Aquila molecular cloud drive $H_2$ outflows with larger opening angles. Does this mean that Class I sources have more likelihood to drive poor collimated $H_2$ outflows? We must note two facts: first, the $H_2$ outflow with the largest opening angle is driven by a Class II source; second, the $H_2$ outflows driven by Class II sources in our sample may be very likely incomplete. \citet{zhang13} estimated that the fraction of Class II sources which drive $H_2$ outflows is $\sim$15\% in Ophiuchus, but this fraction in our sample is $<$2\%. Thus we cannot estimate the collimation of $H_2$ outflows driven by Class II source in the Aquila molecular cloud using our incomplete sample. If we only consider the $H_2$ outflows driven by protostars ($\alpha>$0) in the Aquila molecular cloud, their opening angles show no obvious correlation with the spectral indices of their driving sources (correlation coefficients $<$0.3), being consistent with the studies in other star-forming regions \citep{davis08,davis09,zhang13}.

\citet{davis09} discussed the reasons for the lack of correlations between jet lengths or jet opening angles and spectral indices of driving sources. Firstly, shock-excited $H_2$ emission is not a good tracer of outflow parameters due to its short cooling time. $H_2$ emission is not a sensitive tracer for long jets, which results in the under-estimation of jet lengths. In addition, the wings of jet-driven bow shocks are often wider than the underlying jets and changes in flow direction due to precession, which results in the over-estimation of jet opening angles. Secondly, the precise relationship between source spectral index and source age has not been established yet. 

\subsubsection{Jet position angles}
Previous studies of Orion A \citep{stanke02,davis09}, Perseus \citep{davis08}, and Ophiuchus \citep{zhang13} found that the orientation of outflows shows a homogeneous distribution with no significant trends. However, the study of DR21/W75 by \citet{davis07} found that the molecular hydrogen outflows, in particular from massive cores, are preferentially orthogonal to the molecular ridge, indicating a physical connection between PAs of outflows and the large-scale cloud structure. A histogram of the position angles of molecular hydrogen outflows detected in the Aquila molecular cloud is shown in Fig.~\ref{parahist} (bottom left panel), using a bin size of 30\degr. Note that here we assume all outflows to be bipolar outflows and transfer their position angles to the range of [0\degr, 180\degr]. It seems that there is a peak at 60\degr-90\degr, but the value of the peak is only several counts higher than nearby columns. Actually, a Kolmogorov-Smirnov (KS) test shows that there is a $>$93\% probability that such a distribution is drawn from a homogeneously distributed sample. 
Thus here we propose the distribution of jet position angles to be a homogeneous distribution, which means that outflows in the Aquila molecular cloud may orientate randomly.


\subsubsection{Jet H$_2$ 1-0 S(1) luminosities}
The $H_2~1-0~S(1)$ luminosities of our outflows range from $\sim$3$\times$10$^{-6}$ to $\sim$2.6$\times$10$^{-3}$\,L$_{\sun}$ with the median of 4.7$\times$10$^{-5}$\,L$_{\sun}$ (see table~\ref{outflows_info}). 
{\mybf Figure~\ref{parahist} (bottom right panel) shows the distribution of $H_2~1-0~S(1)$ luminosities of $H_2$ outflows in logarithm (in bins of 0.5). This distribution exhibits a decrease in the number of outflows with increasing $H_2~1-0~S(1)$ luminosities in the range of $log(L_{2.12}/L_{\sun}) > -$4. Note that there is also a decrease in the {\mybfsec number of objects with very low $L_{2.12}$ $H_2$ outflows (log(L$_{2.12} <$ -5))}, which could be due to the incompleteness of low luminosity $H_2$ outflow sample.}

\citet{stanke02} and \citet{ioa2} investigated the outflow luminosities in Orion A and Galactic plane toward Serpens and Aquila, respectively. The $H_2~1-0~S(1)$ luminosities of outflows in Orion A range from $\sim$10$^{-4}$ to $\sim$10$^{-2}$\,L$_{\sun}$ while the luminosities of outflows in the Galactic plane toward Serpens and Aquila range from $\sim$10$^{-3}$ to $\sim$10$^{-1}$\,L$_{\sun}$. They also find that the $H_2~1-0~S(1)$ luminosities of outflows exhibit a power-law behavior. The slopes of power-law functions obtained by them are $-$1.1 in Orion A and $-$1.9 in the Galactic plane toward Serpens and Aquila, respectively. For our sample, the distribution of $H_2~1-0~S(1)$ luminosities is not a well-peaked histogram. The peak of this distribution changes with the histogram bin size. 
Consequently we did not statistically test the distribution of jet $H_2~1-0~S(1)$ luminosities in the Aquila molecular cloud.

\citet{phdstanke} investigated the evolutionary trends in the flow $H_2$ luminosities in Orion A and found that more evolved flow sources seem to drive lower luminosity outflows. \citet{kha12} also found a relatively poor correlation between outflow luminosity in $H_2~1-0~S(1)$ line emission and spectral indices of driving sources based on 13 outflows in Braid Nebula. Figure~\ref{paravsalpha} (right two panels) shows the relation between outflow $H_2~1-0~S(1)$ luminosities and spectral indices of driving sources in the Aquila molecular cloud. We find no obvious correlation (correlation coefficient $<$ 0.2) between $H_2~1-0~S(1)$ luminosities of our outflows and spectral indices of driving sources. The mean $H_2~1-0~S(1)$ luminosity of outflows from Class I sources is 3$\times$10$^{-4}$\,L$_{\sun}$ while the mean $H_2~1-0~S(1)$ luminosity of outflows from Class II/III sources is 1$\times$10$^{-4}$\,L$_{\sun}$ if using de-reddened spectral indices to do the YSO classification. Thus although no clear correlation is found, it seems that there is still a general trend that Class I sources drive more luminous $H_2$ outflows than Class II/III sources. {\mybf Of course, this conclusion is not robust. 
\myeph{At best, it is only a modest observation. However, even if this conclusion  were true, we should like to point our readers to two caveats : (i)} the contribution of $H_2~1-0~S(1)$ line to the $H_2$ luminosity of outflows is only $\sim$10\% \citep{cog06} and, (ii) $H_2$ fluxes are affected by the environment--Class I sources are more likely to be in dense clouds where there is more $H_2$ gas that can be shock excited into
emission while Class II sources are probably more widely distributed in
regions where there isn't very much $H_2$ gas. Therefore, \myeph{any strong suggestion about the evolution of outflow luminosity merely encouraged by observations of the H$_{2}$(1-0) S(1) line-emission is likely to be too far-fetched.}}

\subsection{Momentum injection and cloud support}
\myeph{Feedback from mass outflows injects turbulent energy and momentum into the parent cloud that could possibly support the cloud against self-gravity.} 
\myeph{In this section we discuss the energy budget of the turbulence injected by the detected molecular hydrogen outflows in the Aquila molecular cloud.} \myeph{To this end we calculate the momentum that is likely to be injected by this sample of outflows and investigate if this injected momentum is sufficient to maintain the observed turbulence in the Aquila molecular cloud.}

Following the suggestions by \citet{hhperseus,davis08,ioa2}, we estimate the turbulent momentum in the Aquila molecular cloud using its mass ($M_c$) and velocity dispersion ($\delta V_{tur}$), as $P_{tur}\sim M_c\delta V_{tur}$.

We use our Herschel extinction map obtained in section~\ref{extmap} to estimate the cloud mass of our survey region. Assuming the gas-to-dust ratio of $N(H_2)/A_V=  $0.94$\times$10$^{21}$\,cm$^{-2}$\,mag$^{-1}$ \citep{bohlin78} and a distance of 260\,pc for the Aquila molecular cloud, the total mass in our survey region is about 4200\,M$_{\sun}$. However, the momentum injection from outflows is \myeph{only likely in the relatively dense regions, i.e., those regions having column density above the star-formation threshold for the cloud.} \citet{lada10} found that the star formation rate (SFR) in molecular clouds is linearly proportional to the cloud mass above an extinction threshold of $A_V\approx 7.3$\,mag, corresponding to a gas volume density threshold of n($H_2$) $\approx$ 10$^4$\,cm$^{-3}$. \myeph{Adopting this density as the threshold above which star-formation in the Aquila molecular cloud is likely, we estimate that the mass of potentially star-forming gas in this region is $\sim$2700 M$_{\odot}$.}

\begin{figure*}
\centering
\includegraphics[width=8cm]{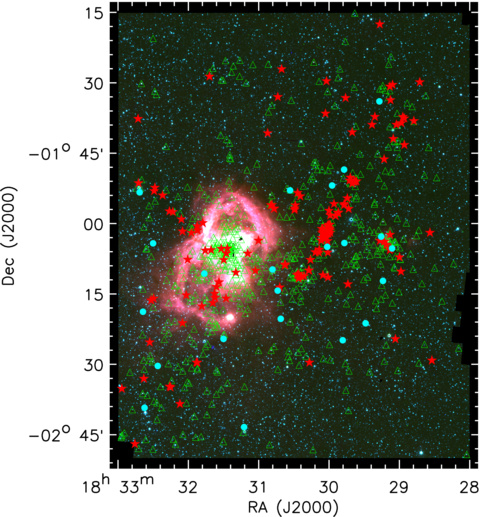}%
\includegraphics[width=8.1cm]{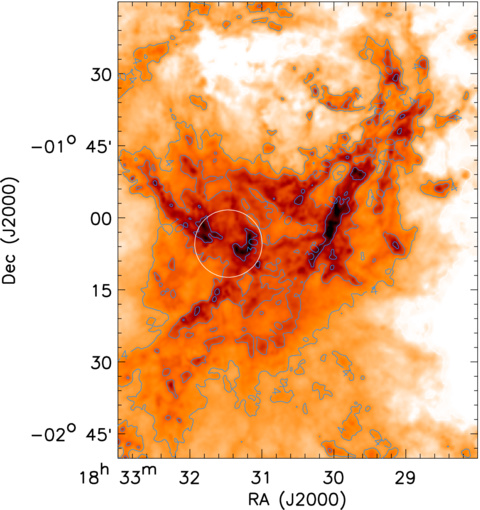}
\caption{{\it Left:} Three-color image of the Aquila molecular cloud that is constructed with IRAC 3.6\,\micron~(blue), 4.5\,\micron~(green), and 8.0\,\micron~(red) images. The YSOs identified in this paper are also marked with different symbols, red pentagrams indicating Class I sources, green triangles indicating Class II sources and cyan circles indicating transition disks. {\it Right:} Extinction map derived from {Herschel} PACS/SPIRE archive data of the Aquila molecular cloud with the resolution of 36.9\arcsec, corresponding to the HPBW resolution of SPIRE at 500\,\micron. The blue contours represent the visual extinction ($A_V$) of 4, 10, and 30 magnitudes. The location of the W40 H {\small II} region is marked with the white circle whose scale is in proportion to the size of W40 \citep{mallick13}.}
\label{spitzercover}
\end{figure*}

\begin{figure*}
\centering
\includegraphics[width=12.0cm]{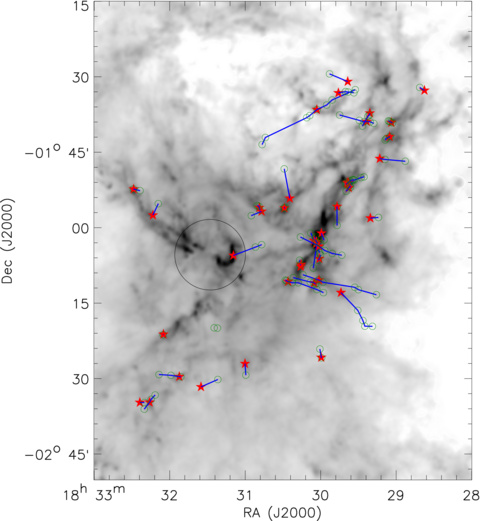}
\caption{Spatial distribution of molecular hydrogen outflows detected in this paper. The underlying is the Herschel extinction map obtained in Section~\ref{extmap}. The 108 MHO features are marked with green circles and 40 outflow sources are labeled with red filled pentagrams. The blue solid lines connect the outflow sources and their associated MHOs, hence indicating the identified molecular hydrogen outflows. The location of W40 H {\small II} region is marked with a black circle whose scale is in proportion to the size of W40 \citep{mallick13}.}
\label{outflows}
\end{figure*}

\begin{figure*}
\centering
\includegraphics[width=7.5cm]{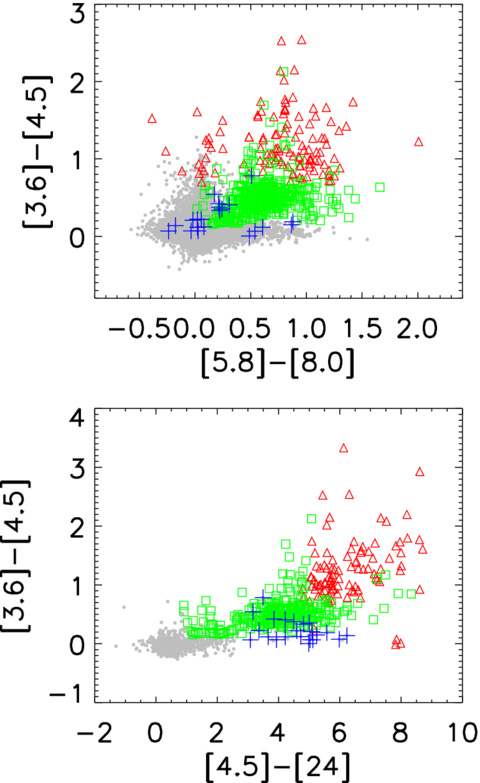}%
\includegraphics[width=6.9cm]{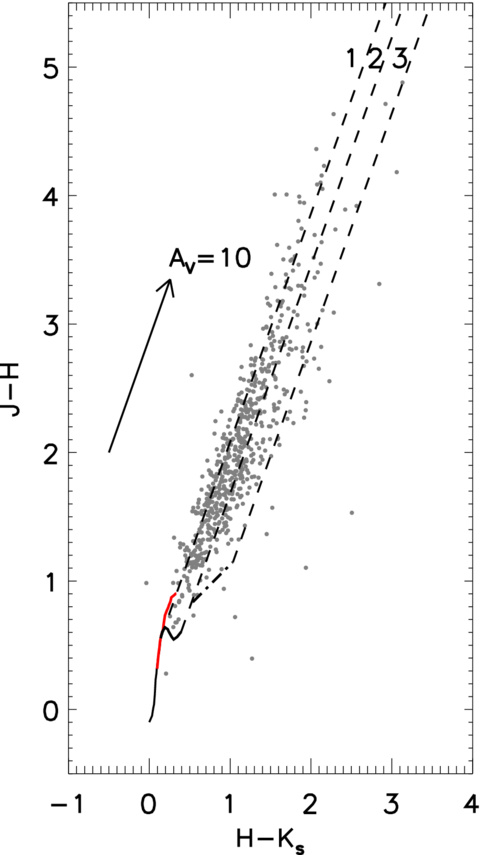}
\caption{{\mybf \textit{Left}: Color-color diagrams showing all classified YSO candidates and field sources. The Class I candidates are marked with red triangles while the Class II candidates are labeled with green squares as well as the transition disk candidates are marked with blue pluses. The grey dots represent the Class III/field sources.} \textit{Right}: The $H-K_s$ vs. $J-H$ color-color diagram for the YSO candidates in the Aquila molecular cloud. The solid line show the intrincsic colors for the main-sequence stars (black) and giants (red) \citep{bb88} while the dash-dotted line is the locus of T Tauri stars from \citet{meyer97}. The dashed lines show the reddening direction and the arrow shows the reddening vector. The extinction law we adopted is from \citet{jiang14}. Note that the dashed lines separate the diagram into three regions marked with numbers 1, 2, and 3 in the figure. In each region, we use different methods to estimate the extinction of the YSO candidates (see the text for details).}
\label{ccdyso}
\end{figure*}

\begin{figure*}
\centering
\includegraphics[width=16cm]{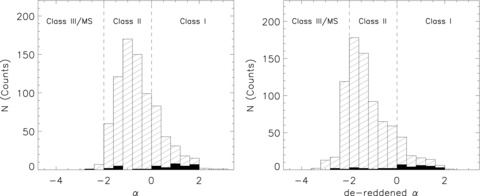}
\caption{Histograms of the apparent spectral indices (left panel) and de-reddened spectral indices (right panel) of all YSOs (hatched columns) and $H_2$ flow sources (filled black columns) identified in this paper. The dashed lines show the criteria of YSO classification scheme suggested by \citet{lada87}.}
\label{yso_alpha}
\end{figure*}

From Fig.~\ref{spitzercover}, we can see that there are mainly two clouds in our survey region, one associated with the Serpens South cluster and another associated with the H {\small II} region, W40. We mark the location of W40 in Fig.~\ref{spitzercover} and Fig.~\ref{outflows} with circles whose scales are in proportion to the size of W40. The size of W40 (7\arcmin) is estimated using radio continuum observation data by \citet{mallick13}. The distance estimation of W40 varies from $\sim$300 to $\sim$900\,pc by different authors \citep{radhak72,vallee87,w40book,shuping12}. Recent study by \citet{shuping12} suggested a distance of $\sim$500\,pc for W40 based on the near-infrared spectral analysis of OB stars in the W40 H {\small II} region, which indicates that W40 may be not associated with Serpens and Serpens South cloud. We also note that, of our 40 outflows, only one may be associated with W40. After excluding the mass of W40 (mass inside the circle marked in Figs.~\ref{spitzercover} and \ref{outflows}), the remain mass above star formation threshold is about 2300\,M$_{\sun}$.

The turbulent velocity is difficult to measure in the molecular cloud. Here we follow the suggestion from \citet{davis08,ioa2} and use the widths of molecular lines to estimate the turbulent motions. The typical value of volume density ($n(H_2)$) for the brightest NH$_3$ emission in dark clouds is $\sim$ 10$^4-$10$^5$\,cm$^{-3}$ \citep{ho83}, indicating that NH$_3$ can be a tracer of dense molecular cloud with star formation activity. \citet{lev13} presented the results of mapping observations in the NH$_3$(1, 1) and (2, 2) lines toward 49 sources in the Aquila region with the Effelsberg 100 m telescope. They detected NH$_3$ emission lines in 19 sources and the widths of lines are from $\sim$0.2 to 1.5\,km\,s$^{-1}$. We adopt a mean value of 0.7\,km\,s$^{-1}$ to estimate turbulent velocity $\delta V_{tur}$. Note that this value is also similar to the mean width of C$^{18}$O (2-1) lines (0.6\,km\,s$^{-1}$) which is used to estimate the turbulent velocity of Perseus by \citet{davis08}. Finally, using the cloud mass and turbulent velocity we can obtain the turbulent momentum of $\sim$1600\,M$_{\sun}$\,km\,s$^{-1}$ in our survey region.

\myeph{Since the momentum of an outflow calculated using only a single emission line at 2.12\,\micron~is unlikely to be accurate, we also adopt canonical values suggested by \citet{davis08} and \citet{ioa2} to estimate the momentum injected by these outflows. For typical magnitudes of outflow momentum in the range  $\sim10^{-4}$ and $\sim2\times 10^{-6}$ M$_{\sun}$\,km\,s$^{-1}$\,yr$^{-1}$, for Class 0 and Class I sources, respectively \citep{bon96}, we estimate that a low-mass protostar will only inject $\sim$1.0\,M$_{\sun}$\,km\,s$^{-1}$ \citep{davis08,ioa2}, over its lifetime that is usually on the order of 10$^{4}$$-$10${^5}$ yrs. Our calculations therefore suggest, the sample of outflows detected in Aquila can only possibly supply $\sim$40\,M$_{\sun}$\,km\,s$^{-1}$, which is a factor of 40 smaller than the observationally estimated cloud turbulent momentum. Evidently, the molecular hydrogen outflows appear to be inefficient sources of turbulent momentum for this cloud.}

\myeph{However, a caveat must be added here; the canonical values for outflow momentum suggested by \citet{bon96} were originally calculated using measurements of ``classical" slow CO molecular outflows. These suggested values of momentum therefore only represent the low-velocity components of mass outflows.} However, a large fraction of the jet momentum may be carried by a high-velocity collimated jet \citep{hhperseus}. \myeph{It is therefore likely that we are underestimating the momentum injected by the MHOs in the Aquila molecular cloud.} Nevertheless, it has been argued that the transport of the fast jet momentum to the ambient cloud may be inefficient because the jets may be relatively unhindered by their surroundings, especially considering the existence of parsec-scale jets and the high proper motions of the distant jet knots \citep{davis08}. 

One possible way to balance the observed turbulent momentum of the molecular cloud is via many generations of outflows, as suggested by \citet{davis08} in Perseus. The lifetime of the giant molecular clouds is $\sim$10$^7$\,yr \citep{larson81}, which is one or two orders of magnitude longer than the lifetime of Class 0/I phase. Thus a few tens of generations of mass outflows may contribute to the observed turbulent momentum of the molecular clouds.

\section{Summary}\label{summary}
We have performed an unbiased near-infrared survey toward the Aquila molecular cloud in $J$, $H$, $K_s$, and $H_2$ bands using WIRCam equipped on CFHT. Our survey covers a sky region of $\sim$ 1\degr$\times$1\degr~which embraces the Serpens South cluster and W40 H {\small II} region. Combining the archival data from {\it Spitzer} and {Herschel} space telescopes, we investigate the properties of molecular hydrogen outflows in our survey region. The main results are listed as follows:

\begin{itemize}
\item[1.] We have identified 45 MHOs that consist of 108 MHO features. Of 45 MHOs, 11 are previously known objects and 34 are newly discovered. Using the {\it Spitzer} archival data, we also identify 802 YSO candidates in our survey region. Based on the morphologies of MHOs and locations of MHOs and YSO candidates, we associate 43 MHOs with 40 YSO candidates and obtain 40 molecular hydrogen outflows.

\item[2.] We obtain the visual extinction map of our survey region using the {Herschel} archival data through fitting the SED from 160-500\,\micron~pixel by pixel. Based on this extinction map, we have tried to correct the fluxes of YSO candidates for extinction and classify 802 YSO candidates into three classes with the de-reddened spectral indices: {\mybf 100} Class I sources, {\mybf 551} Class II sources, and {\mybf 151} Class III/MS sources. We find that 27 ($\sim$68\%) of 40 outflow sources are protostars while $\sim${\mybf 27\%}~of Class I sources drive $H_2$ outflows. We also find that 18 outflow sources are associated with infrared nebulae.

\item[3.] In our outflow sample, the distribution of jet lengths shows a steep decrease in the number of outflows with increasing lengths, roughly following the relation of $N \propto 10^{(-0.85\pm0.1)\times length(pc)}$. The slope of $-$0.85 is similar to the value of $-$0.75 that is obtained with outflows from the low/intermediate mass stars in the Galactic plane toward Serpens and Aquila by \citet{ioa2}. The distribution of jet position angles shows a nearly uniform distribution, which indicates that the molecular hydrogen outflows in the Aquila molecular cloud may orientate randomly. We also find that there is no obvious correlation between jet lengths or jet opening angles and spectral indices of driving sources, which agrees with the results in Perseus \citep{davis08}, Orion A \citep{davis09}, and Ophiuchus \citep{zhang13}.

\item[4.] We perform the areal photometry for our MHO features and estimate the $H_2~1-0~S(1)$ line luminosities for the 40 outflows in the Aquila molecular cloud. The $H_2~1-0~S(1)$ luminosities of our outflows range from $\sim$3$\times$10$^{-6}$ to $\sim$2.6$\times$10$^{-3}$ L$_{\sun}$ with the median value of 4.7$\times$10$^{-5}$ L$_{\sun}$, which is one or two order of magnitude lower than that in Orion A \citep{stanke02} and that in the Galactic plane toward Serpens and Aquila \citep{ioa1,ioa2}. We also find no obvious correlation between outflow $H_2~1-0~S(1)$ luminosities and spectral indices of driving sources.


\item[5.] We estimate that the turbulent momentum in our survey region is about 1600 M$_{\sun}$\,km\,s$^{-1}$. However, the momentum injection from molecular hydrogen outflows in our survey region is only $\sim$40 M$_{\sun}$\,km\,s$^{-1}$, which is a factor of 40 smaller than the cloud turbulent momentum. Thus the momentum injection from molecular hydrogen outflows is not enough to match the observed turbulent momentum in our survey region of the Aquila molecular cloud.

\end{itemize}



\acknowledgments
The authors acknowledge Shaobo Zhang and Yao Liu for very valuable comments and discussions. We thank Shaobo Zhang for his IDL plotting code which has been uploaded to the web site\footnote{\url{http://code.google.com/p/aicer/}}. We acknowledges the support by NSFC grants 11173060, 11233007, and 11127903. This work is supported by the Strategic Priority Research Program ``The Emergence of Cosmological Structures" of the Chinese Academy of Sciences, grant No. XDB09000000. {\mybfsec Sumedh Anathpindika is partially supported by the Young scientist grant YSS/2014/000304 awarded by the Department of Science \& Technology, Government of India.} This work is based on observations obtained with WIRCam, a joint project of CFHT, Taiwan, Korea, Canada, France, at the Canada-France-Hawaii Telescope (CFHT) which is operated by the National Research Council (NRC) of Canada, the Institute National des Sciences de l'Univers of the Centre National de la Recherche Scientifique of France, and the University of Hawaii. This work is based on observations made with the Spitzer Space Telescope, which is operated by the Jet Propulsion Laboratory, California Institute of Technology under a contract with NASA. This research has made use of data from the Herschel Gould Belt survey (HGBS) project (\url{http://gouldbelt-herschel.cea.fr}). The HGBS is a Herschel Key Programme jointly carried out by SPIRE Specialist Astronomy Group 3 (SAG 3), scientists of several institutes in the PACS Consortium (CEA Saclay, INAF-IFSI Rome and INAF-Arcetri, KU Leuven, MPIA Heidelberg), and scientists of the Herschel Science Center (HSC). {\mybfsec This research uses data obtained through the Telescope Access Program (TAP), which has been funded by the National Astronomical Observatories of China, the Chinese Academy of Sciences (the Strategic Priority Research Program ``The Emergence of Cosmological Structures" Grant No. XDB09000000), and the Special Fund for Astronomy from the Ministry of Finance.} This research has also made use of the SIMBAD database, operated at CDS, Strasbourg, France. In particular, the Aladin image server is very useful in locating previously known objects on our frames.






\bibliographystyle{apj}
\bibliography{myref}{}
\clearpage



\begin{deluxetable}{cccccclccccc}
\rotate
\centering
\tabletypesize{\tiny}
\tablewidth{0pt}
\tablecaption{Molecular hydrogen outflows identified in Aquila.\label{outflows_info}}
\tablehead{
\colhead{Outflow source}&
\colhead{RA\tablenotemark{a}}&
\colhead{Dec\tablenotemark{a}}&
\colhead{$\alpha$\tablenotemark{a}}&
\colhead{$\alpha$\arcmin\tablenotemark{b}}&
\colhead{A$_V$\tablenotemark{c}}&
\colhead{MHO\tablenotemark{d}}&
\colhead{L\tablenotemark{e}}&
\colhead{$\theta$\tablenotemark{f}}&
\colhead{PA\tablenotemark{g}}&
\colhead{$H_2~1-0~S(1)$ Luminosity\tablenotemark{h}}&
\colhead{Outflow}\\
\colhead{ID}&
\colhead{(J2000)}&
\colhead{(J2000)}&
\colhead{}&
\colhead{}&
\colhead{(mag)}&
\colhead{}&
\colhead{(pc)}&
\colhead{(deg)}&
\colhead{(deg)}&
\colhead{(10$^{-5}$ L$_{\sun}$)}&
\colhead{type}
}
\startdata
1 & 18 28 37.6  &-01 32 43&-1.7&-1.8&0.6&3259&0.09&8  &59 &4.07$\pm$1.55&unipolar\\
2 & 18 29 03.9  &-01 39 06&\ldots&\ldots&7.0&3260&0.08&62 &57 &8.42$\pm$2.86&bipolar\\
3 & 18 29 05.3  &-01 41 56&0.9&0.8&11.9&2213&0.11&66 &118&14.80$\pm$4.96&bipolar\\
4 & 18 29 13.1  &-01 46 17&0.3&0.3&5.7&3261&0.39&6  &261&1.46$\pm$0.76&unipolar\\
5 & 18 29 20.5  &-01 58 07&-2.4&-2.5&5.9&3262&0.13&3  &275&0.40$\pm$0.24&unipolar\\
6 & 18 29 20.9  &-01 37 14&0.8&-0.3&25.9&3263&0.23&34 &160&1.75$\pm$0.79&unipolar\\
7 & 18 29 23.4  &-01 38 55&1.6&1.4&4.7&3264&0.53&32 &86 &9.80$\pm$3.98&bipolar\\
8 \tablenotemark{i}& 18 29 37.6  &-01 52 05&-1.4&-1.5&13.1&3265&0.24&29 &138&3.89$\pm$1.50&bipolar\\
9 \tablenotemark{i}& 18 29 38.1  &-01 51 01&1.8&1.4&18.0&2214&0.31&39 &128&38.30$\pm$13.50&bipolar\\
10& 18 29 38.5  &-01 30 58&-0.7&-1.1&9.2&3266&0.33&6  &68 &43.10$\pm$24.00&unipolar\\
11\tablenotemark{i}& 18 29 38.9  &-01 51 07&1.9&1.0&12.9&3267&0.10&65 &68 &1.88$\pm$0.73&bipolar\\
12& 18 29 43.9  &-02 12 55&2.0&1.9&5.8&3268&0.71&11 &219&66.40$\pm$26.20&unipolar\\
13& 18 29 45.9  &-01 33 13&0.0&0.0&1.4&3269&0.24&12 &275&5.66$\pm$2.51&unipolar\\
14& 18 29 47.0  &-01 55 48&1.0&0.5&13.3&3270&0.29&64 &176&4.57$\pm$1.56&unipolar\\
15& 18 29 59.5  &-02 01 07&0.6&0.4&29.1&3271&0.12&7  &201&0.70$\pm$0.34&unipolar\\
16& 18 29 59.6  &-02 25 49&-1.4&-1.7&4.2&3272&0.14&149&44 &25.20$\pm$9.66&bipolar\\
17\tablenotemark{i}& 18 30 01.0  &-02 06 10&0.9&0.0&12.1&3251&0.10&32 &157&2.60$\pm$1.16&bipolar\\
18\tablenotemark{i}& 18 30 01.3  &-02 03 43&1.9&0.8&24.0&3250&0.69&22 &58 &27.00$\pm$10.90&bipolar\\
19& 18 30 01.4  &-02 10 26&\ldots&\ldots&11.9&3252&1.23&111&73 &111.00$\pm$42.00&bipolar\\
20& 18 30 03.1  &-01 36 33&1.5&1.5&3.6&3273&1.65&90 &126&33.80$\pm$13.60&bipolar\\
21& 18 30 03.4  &-02 02 46&1.2&0.2&37.2&3247&0.12&134&46 &75.30$\pm$29.10&bipolar\\
22\tablenotemark{i}& 18 30 03.5  &-02 03 10&1.8&1.7&33.7&3248;3251b2&0.51&78 &175&258.00$\pm$101.00&bipolar\\
23& 18 30 05.2  &-02 11 00&0.3&0.2&8.4&3254a;3255&0.36&143&65 &74.20$\pm$30.70&bipolar\\
24\tablenotemark{i}& 18 30 05.3  &-02 02 34&0.2&-0.7&27.3&3249&0.34&8  &23 &18.00$\pm$7.51&bipolar\\
25& 18 30 15.6  &-02 07 20&1.6&1.5&18.7&3274&0.02&22 &119&0.42$\pm$0.21&unipolar\\
26& 18 30 15.9  &-02 07 43&0.2&-0.7&18.8&3275&0.10&3  &10 &0.30$\pm$0.15&unipolar\\
27& 18 30 24.5  &-01 54 11&1.0&0.2&20.7&3276&0.46&1  &11 &2.87$\pm$1.49&unipolar\\
28& 18 30 25.9  &-02 10 43&1.6&1.4&9.3&3253;3254b;3254c;3277;3278&0.61&31 &71 &19.50$\pm$7.08&bipolar\\
29& 18 30 28.8  &-01 56 06&1.1&1.0&15.2&3279&0.02&106&97 &1.15$\pm$0.57&bipolar\\
30& 18 30 46.9  &-01 56 46&1.8&1.7&8.3&3280&0.17&3  &113&2.97$\pm$1.20&unipolar\\
31& 18 30 48.7  &-01 56 02&1.5&1.2&8.9&3281&0.09&49 &32 &3.18$\pm$1.17&bipolar\\
32& 18 31 00.1  &-02 27 02&-1.6&-1.6&6.2&3282&0.18&5  &184&3.43$\pm$1.42&unipolar\\
33& 18 31 09.3  &-02 05 32&-1.3&-2.5&22.6&3283&0.48&5  &291&27.40$\pm$11.60&unipolar\\
34& 18 31 35.1  &-02 31 39&-1.8&-2.2&11.4&3286&0.28&1  &294&2.23$\pm$1.00&unipolar\\
35& 18 31 52.3  &-02 29 38&0.1&0.1&5.8&3287&0.36&27 &85 &7.49$\pm$2.82&bipolar\\
36& 18 32 04.9  &-02 21 12&1.1&1.0&8.7&3288&0.01&36 &151&0.34$\pm$0.15&unipolar\\
37& 18 32 13.2  &-01 57 31&1.2&1.1&9.2&3289&0.20&3  &335&2.14$\pm$0.80&unipolar\\
38& 18 32 16.0  &-02 34 43&0.5&0.3&5.7&3290&0.29&52 &140&16.90$\pm$5.95&bipolar\\
39& 18 32 23.7  &-02 34 45&-1.3&-1.9&7.4&3291&0.07&4  &269&0.68$\pm$0.29&unipolar\\
40& 18 32 28.5  &-01 52 20&1.2&1.2&7.7&3292&0.11&90 &61 &4.73$\pm$1.75&bipolar\\
\enddata
\tablenotetext{a}{The locations, and apparent spectral indices of the outflow sources.}
\tablenotetext{b}{The de-reddened spectral indices of the outflow sources.}
\tablenotetext{c}{The visual extinction values that are used for de-reddening.}
\tablenotetext{d}{The associated MHOs.}
\tablenotetext{e}{Jet lenghts.}
\tablenotetext{f}{Jet opening angles.}
\tablenotetext{g}{Jet position angles. Note that we select $<$180\arcdeg~position angles for the bipolar outflows.}
\tablenotetext{h}{$H_2~1-0~S(1)$ line luminostiy of the outflows.}
\tablenotetext{i}{The parameters of these outflows have large uncertaintities because of the overlaps or interactions between these outflows and ambient other outflows.}
\end{deluxetable}

\clearpage

\begin{figure*}
\centering
\includegraphics[width=8.0cm]{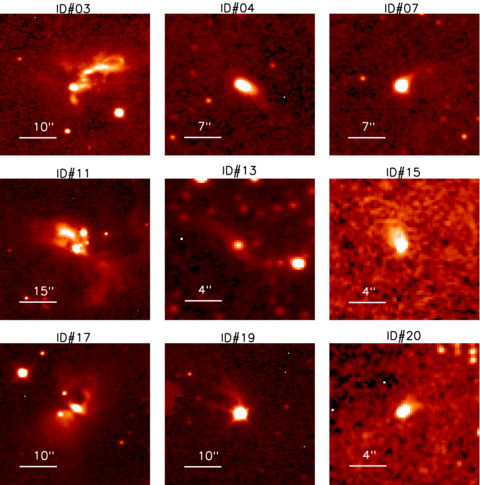}%
\vspace{0.5cm}
\includegraphics[width=8.0cm]{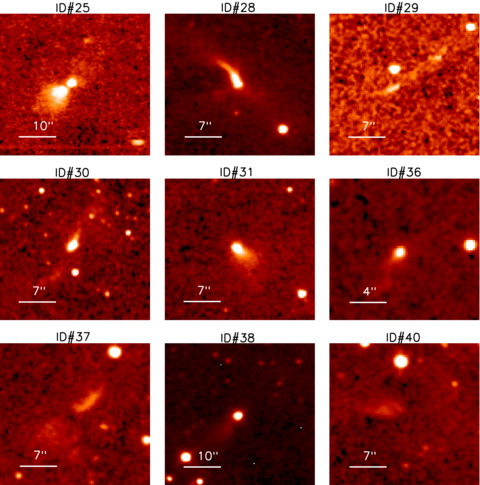}
\caption{Infrared nebulae associated with outflow sources. The backgrounds are $K_s-$band images. North is up and east to the left. {\mybf The outflow sources are located in the center of panels.}}
\label{nebshow}
\end{figure*}

\begin{figure*}
\centering
\includegraphics[width=14cm]{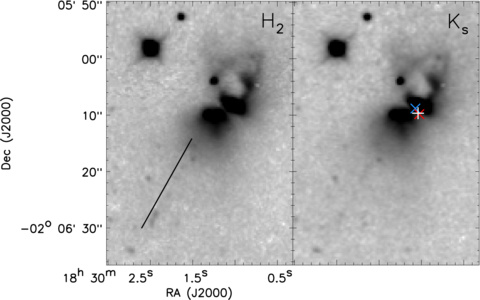}
\caption{A disk-jet system candidate that consists of MHO 3251b1 and its driving source ID\#17. The \textit{left} panel shows the $H_2$ narrow-band image while the \textit{right} panel shows the $K_s$ broadband image. The white plus in the right panel marks the flux-weighted mean IRAC and MIPS 24\,\micron~position for ID\#17. The blue cross in the right panel marks the mean IRAC position for ID\#17 while the red cross labels the MIPS 24\,\micron~position for ID\#17. The solid line in the left panel shows the orientation of a collimated jet.}
\label{jet-disk}
\end{figure*}
\clearpage

\begin{figure*}
\centering
\includegraphics[width=16cm]{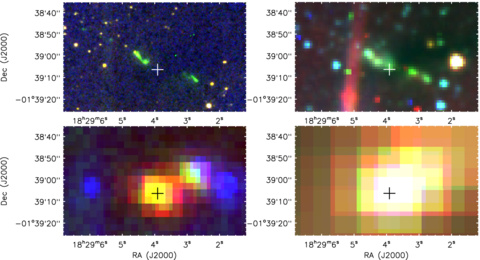}
\caption{The region of MHO 3260. \textit{Top-left}: three-color image constructed with CFHT $J$ (blue), $H_2$ (green), and $K_s$ (red) images. \textit{Top-right}: three-color image constructed with IRAC 3.6\,\micron~(blue), 4.5\,\micron~(green), and 8.0\,\micron~(red) images. \textit{Bottom-left}: three-color image constructed with MIPS 24\,\micron~(blue), PACS 70\,\micron~(green), and 160\,\micron~(red) images. \textit{Bottom-right}: three-color image that is constructed with SPIRE 250\,\micron~(blue), 350\,\micron~(green), and 500\,\micron~(red) images. The plus marks the position of Class 0 source, ID\#2.}
\label{jet-protostar}
\end{figure*}

\begin{figure*}
\centering
\includegraphics[width=14.0cm]{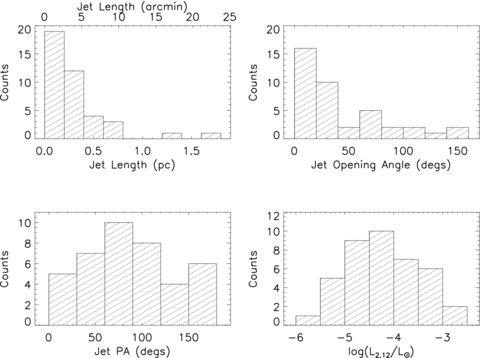}
\caption{Distributions of jet lengths in bins of 0.2\,pc ({\it top left}), jet opening angles in bins of 20\degr~({\it top right}), jet position angles in bins of 30\degr~({\it bottom left}), and logarithm of jet $H_2~1-0~S(1)$ luminosity in bins of 0.5 ({\it bottom right}).}
\label{parahist}
\end{figure*}

\begin{figure*}
\centering
\includegraphics[width=8cm]{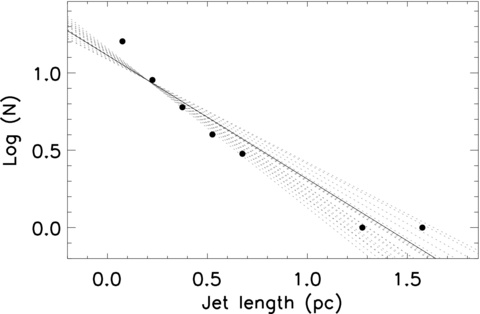}
\caption{Distribution of jet lengths. The black filled circles show the distribution of jet lengths in bins of 0.15\,pc while the solid line shows the linear fitting for this distribution. The gray dotted lines show the linear fittings for the distributions of jet lengths in different bins (0.1-0.3\,pc in step of 0.01\,pc) that have been scaled to the fitting in bins of 0.15\,pc.}
\label{jetlength}
\end{figure*}

\begin{figure*}
\centering
\includegraphics[width=15.0cm]{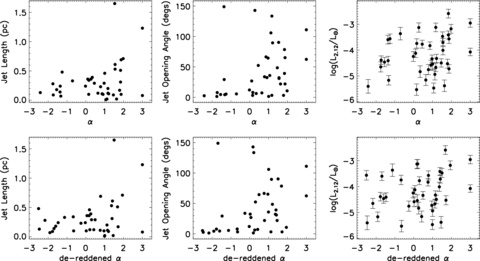}
\caption{Jet length ({\it left}), jet opening angle ({\it middle}), and logarithm of jet $H_2~1-0~S(1)$ luminosity ({\it right}) plot against source apparent spectral index ({\it top}) and source de-reddened spectral index ({\it bottom}). Note that outflow source ID\#2 is invisible at 2-24\,\micron~and ID\#19 is saturated in all IRAC bands and MIPS 24\micron~band. To plot these sources on the figure, we assume the value of 3 for their spectral indices because they are associated with the protostars identified by \cite{bontemps10}.}
\label{paravsalpha}
\end{figure*}

\clearpage
\appendix
\setcounter{figure}{0} \renewcommand{\thefigure}{A\arabic{figure}} 
\setcounter{table}{0} \renewcommand{\thetable}{A\arabic{table}} 

\section{Description of the H$_2$ outflows in Aquila}
\begin{figure}
\includegraphics[width=14cm]{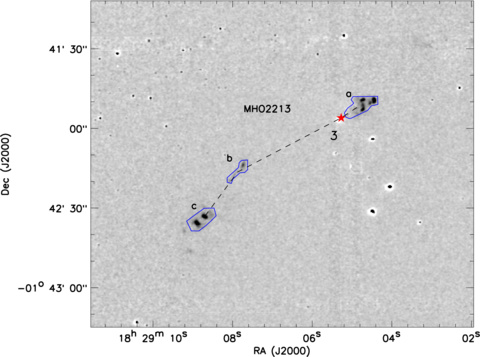}
\caption{Region of MHO 2213. The background is the continuum-subtracted $H_2$ image and the MHO features are marked with blue and pink polygons. The black dashed line connects the MHO features that belongs to the same molecular hydrogen outflow and their possible driving source that is labeled with red filled pentagram. The number marked around the pentagram corresponds to the ID in column 1 of table~\ref{outflows_info}.}
\label{figa01}
\end{figure}

\begin{figure}
\includegraphics[width=14cm]{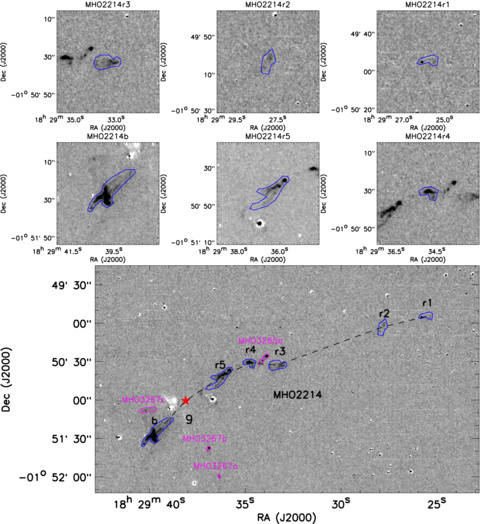}
\caption{Same as Fig.~\ref{figa01}, but for the region of MHO 2214. Note that the square images titled with the names of MHO features show the zoomed views of MHO features which belong to the same $H_2$ outflow.}
\label{figa02}
\end{figure}

\begin{figure}
\includegraphics[width=14cm]{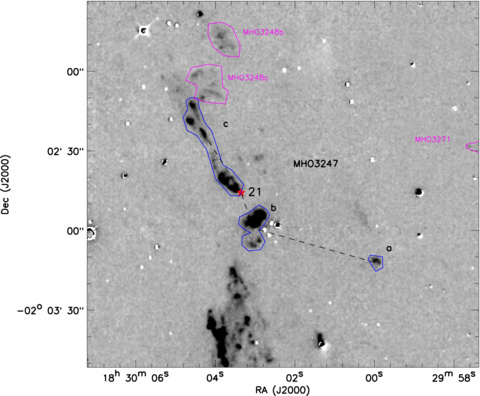}
\caption{Same as Fig.~\ref{figa01}, but for the region of MHO 3247.}
\label{figa03}
\end{figure}

\begin{landscape}
\begin{figure}
\includegraphics[width=20cm]{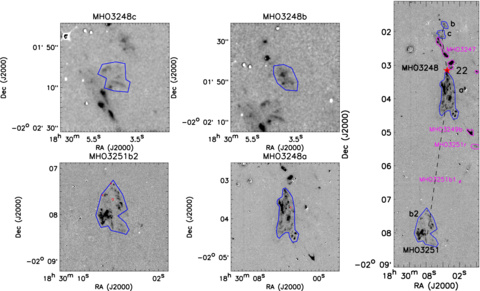}
\caption{Same as Fig.~\ref{figa02}, but for the region of MHO 3248 and MHO 3251b2. Note that the red circles mark the holes inside polygons.}
\label{figa04}
\end{figure}
\end{landscape}

\begin{figure}
\includegraphics[width=14cm]{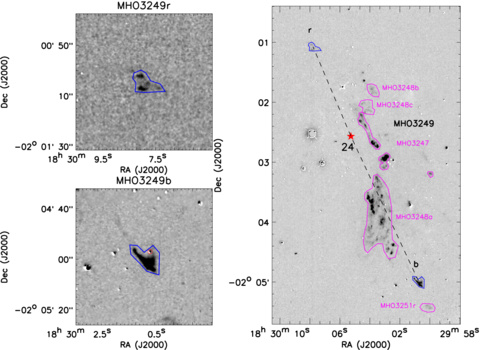}
\caption{Same as Fig.~\ref{figa02}, but for the region of MHO 3249. Note that the red circle marks the hole inside the polygon.}
\label{figa05}
\end{figure}

\begin{figure}
\includegraphics[width=14cm]{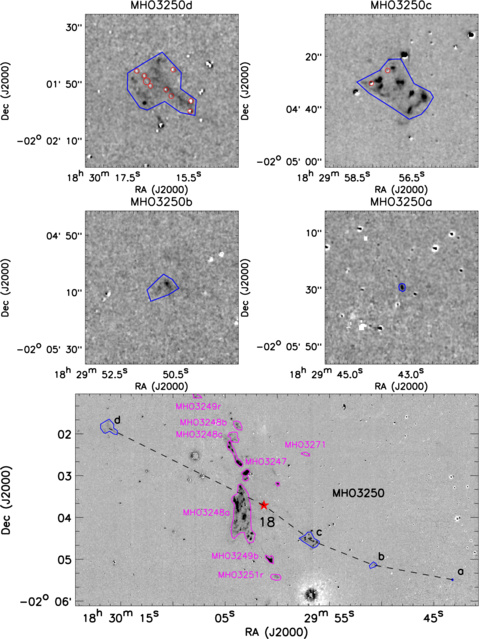}
\caption{Same as Fig.~\ref{figa02}, but for the region of MHO 3250. Note that the red circles mark the holes inside the polygons.}
\label{figa06}
\end{figure}

\begin{figure}
\includegraphics[width=14cm]{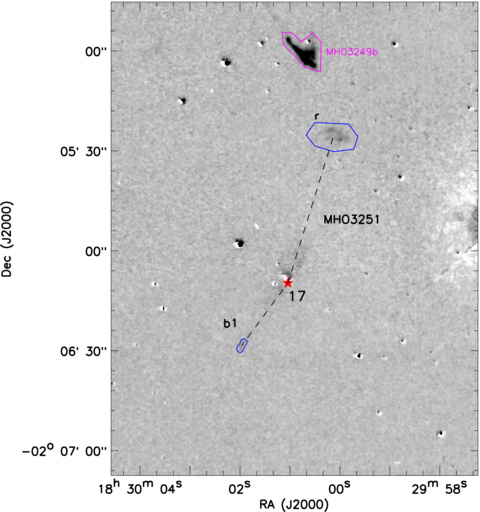}
\caption{Same as Fig.~\ref{figa01}, but for the region of MHO 3251.}
\label{figa07}
\end{figure}

\begin{figure}
\includegraphics[width=14cm]{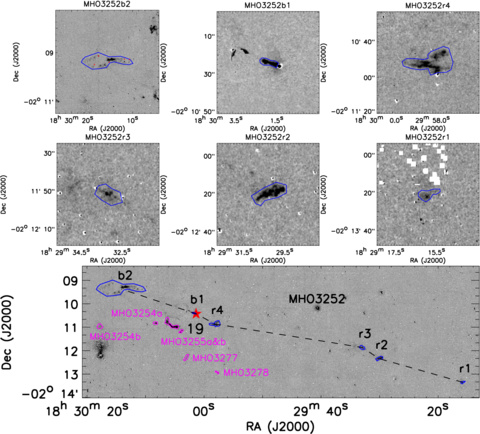}
\caption{Same as Fig.~\ref{figa02}, but for the region of MHO 3252. Note that the red circles mark the holes inside the polygons.}
\label{figa08}
\end{figure}

\begin{figure}
\includegraphics[width=14cm]{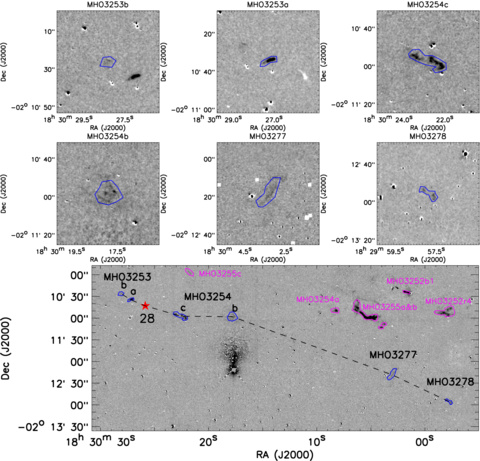}
\caption{Same as Fig.~\ref{figa02}, but for the region of MHO 3253, MHO 3254b, and MHO 3277-3278.}
\label{figa09}
\end{figure}

\begin{figure}
\includegraphics[width=14cm]{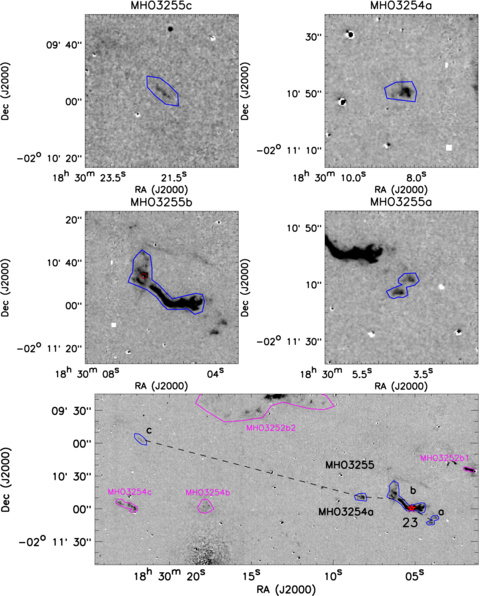}
\caption{Same as Fig.~\ref{figa02}, but for the region of MHO 3254a and MHO 3255. Note that the red circle marks the hole inside the polygon.}
\label{figa10}
\end{figure}

\begin{figure}
\includegraphics[width=14cm]{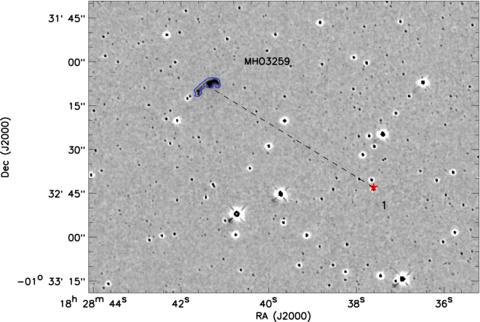}
\caption{Same as Fig.~\ref{figa01}, but for the region of MHO 3259.}
\label{figa11}
\end{figure}

\begin{landscape}
\begin{figure}
\includegraphics[width=20cm]{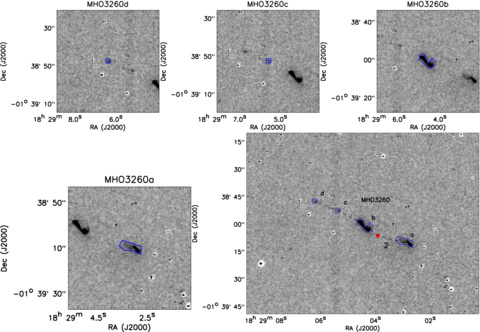}
\caption{Same as Fig.~\ref{figa02}, but for the region of MHO 3260.}
\label{figa12}
\end{figure}
\end{landscape}

\begin{figure}
\includegraphics[width=14cm]{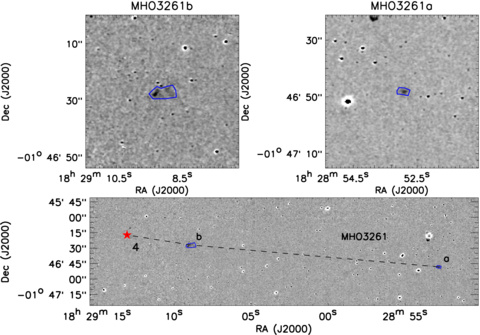}
\caption{Same as Fig.~\ref{figa02}, but for the region of MHO 3261.}
\label{figa13}
\end{figure}

\begin{figure}
\includegraphics[width=14cm]{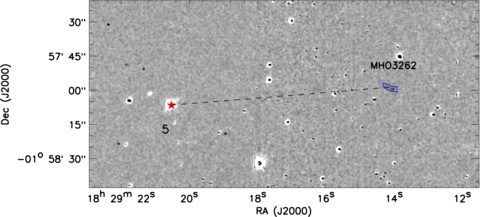}
\caption{Same as Fig.~\ref{figa01}, but for the region of MHO 3262.}
\label{figa14}
\end{figure}

\begin{figure}
\includegraphics[width=14cm]{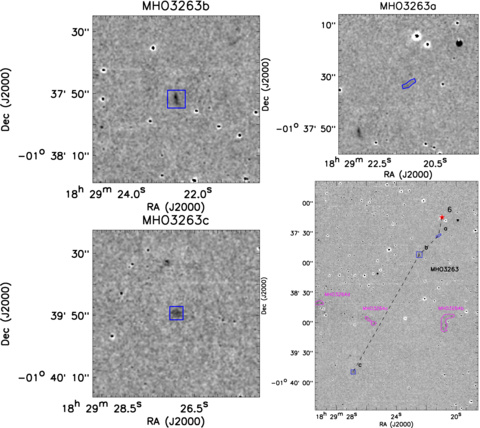}
\caption{Same as Fig.~\ref{figa02}, but for the region of MHO 3263.}
\label{figa15}
\end{figure}

\begin{landscape}
\begin{figure}
\includegraphics[width=20cm]{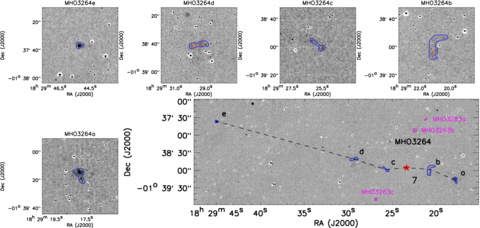}
\caption{Same as Fig.~\ref{figa02}, but for the region of MHO 3264. Note that the red circles mark the holes inside polygons.}
\label{figa16}
\end{figure}
\end{landscape}

\begin{figure}
\includegraphics[width=14cm]{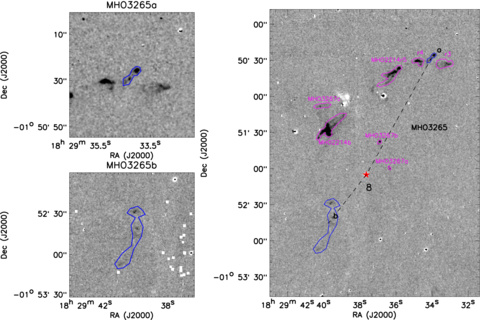}
\caption{Same as Fig.~\ref{figa02}, but for the region of MHO 3265.}
\label{figa17}
\end{figure}

\begin{figure}
\includegraphics[width=14cm]{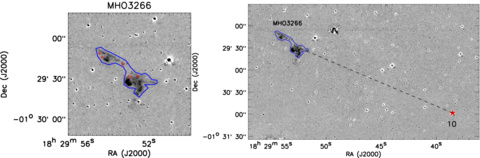}
\caption{Same as Fig.~\ref{figa01}, but for the region of MHO 3266. Note that the red circles mark the holes inside the polygons.}
\label{figa18}
\end{figure}

\begin{figure}
\includegraphics[width=14cm]{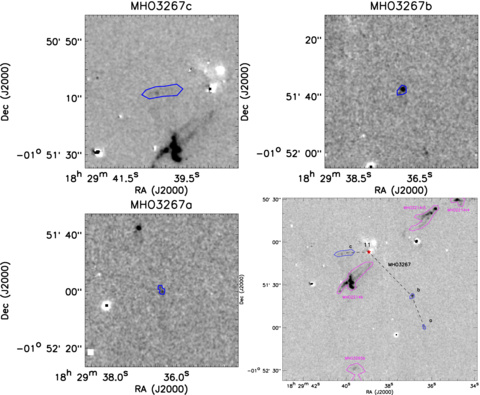}
\caption{Same as Fig.~\ref{figa02}, but for the region of MHO 3267.}
\label{figa19}
\end{figure}

\begin{landscape}
\begin{figure}
\includegraphics[width=20cm]{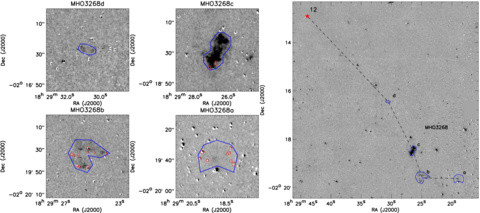}
\caption{Same as Fig.~\ref{figa02}, but for the region of MHO 3268. Note that the red circles mark the holes inside the polygons.}
\label{figa20}
\end{figure}
\end{landscape}

\begin{landscape}
\begin{figure}
\includegraphics[width=20cm]{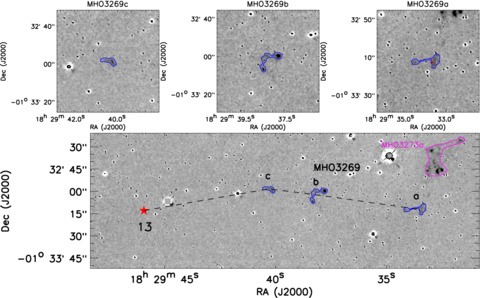}
\caption{Same as Fig.~\ref{figa02}, but for the region of MHO 3269. Note that the red circle marks the hole inside the polygon.}
\label{figa21}
\end{figure}
\end{landscape}

\begin{figure}
\includegraphics[width=14cm]{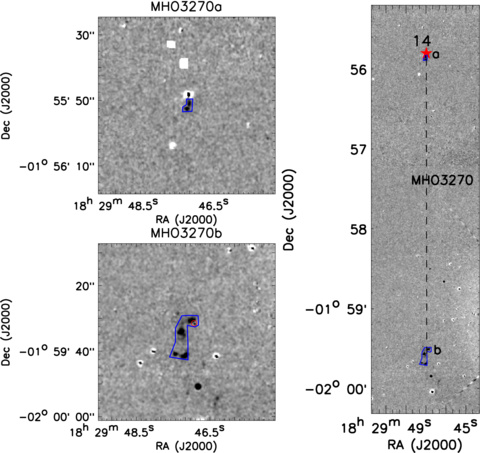}
\caption{Same as Fig.~\ref{figa02}, but for the region of MHO 3270. Note that the red circle marks the hole inside the polygon.}
\label{figa22}
\end{figure}

\begin{landscape}
\begin{figure}
\includegraphics[width=20cm]{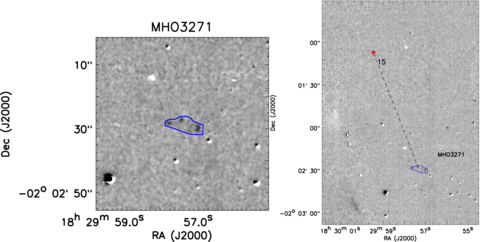}
\caption{Same as Fig.~\ref{figa02}, but for the region of MHO 3271.}
\label{figa23}
\end{figure}
\end{landscape}

\begin{figure}
\includegraphics[width=14cm]{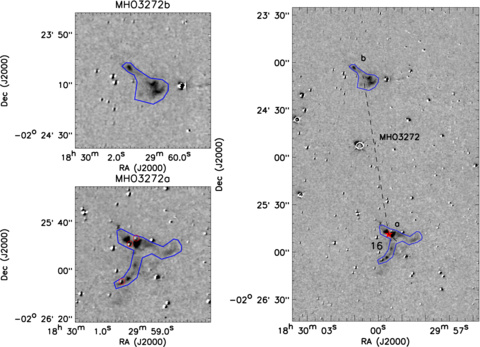}
\caption{Same as Fig.~\ref{figa01}, but for the region of MHO 3272. Note that the red circles mark the holes inside the polygons.}
\label{figa24}
\end{figure}

\begin{figure}
\includegraphics[width=14cm]{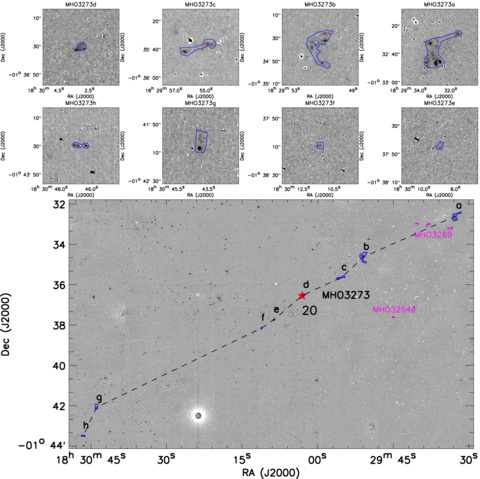}
\caption{Same as Fig.~\ref{figa02}, but for the region of MHO 3273. Note that the red circles mark the holes inside the polygons.}
\label{figa25}
\end{figure}

\begin{figure}
\includegraphics[width=14cm]{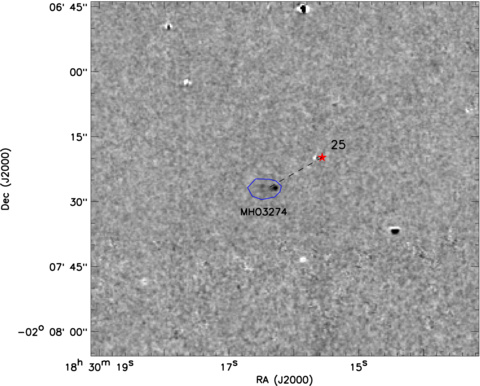}
\caption{Same as Fig.~\ref{figa01}, but for the region of MHO 3274.}
\label{figa26}
\end{figure}

\begin{figure}
\includegraphics[width=14cm]{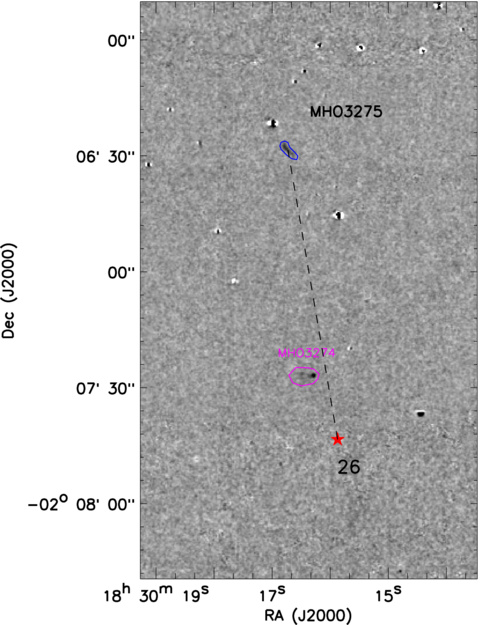}
\caption{Same as Fig.~\ref{figa01}, but for the region of MHO 3275.}
\label{figa27}
\end{figure}

\begin{landscape}
\begin{figure}
\includegraphics[width=20cm]{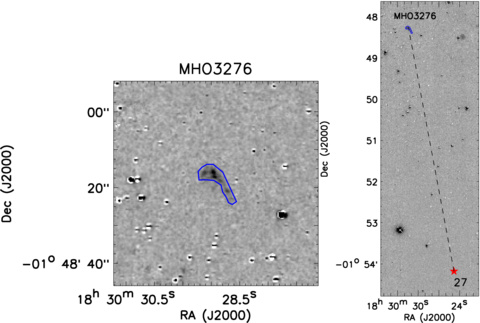}
\caption{Same as Fig.~\ref{figa02}, but for the region of MHO 3276.}
\label{figa28}
\end{figure}
\end{landscape}

\begin{figure}
\includegraphics[width=14cm]{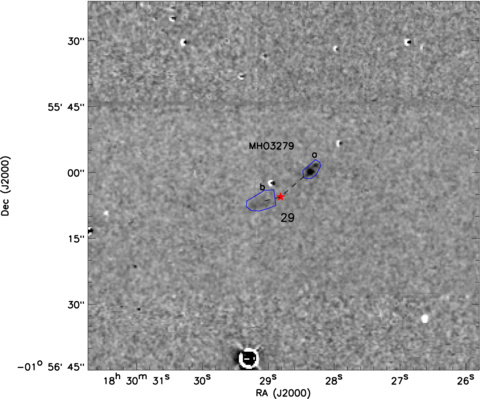}
\caption{Same as Fig.~\ref{figa01}, but for the region of MHO 3279.}
\label{figa29}
\end{figure}

\begin{figure}
\includegraphics[width=14cm]{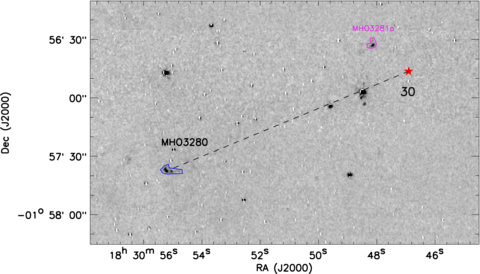}
\caption{Same as Fig.~\ref{figa01}, but for the region of MHO 3280.}
\label{figa30}
\end{figure}

\begin{figure}
\includegraphics[width=14cm]{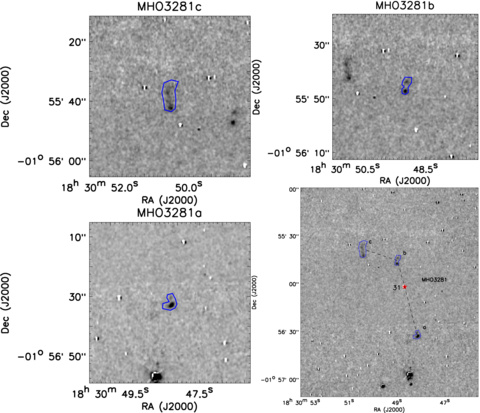}
\caption{Same as Fig.~\ref{figa02}, but for the region of MHO 3281.}
\label{figa31}
\end{figure}

\begin{landscape}
\begin{figure}
\includegraphics[width=20cm]{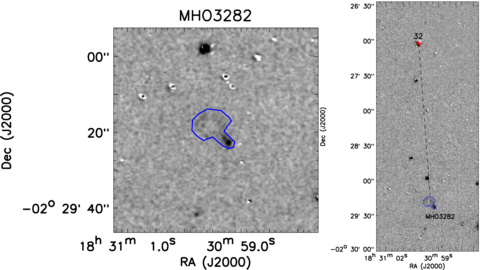}
\caption{Same as Fig.~\ref{figa02}, but for the region of MHO 3282.}
\label{figa32}
\end{figure}
\end{landscape}

\begin{figure}
\includegraphics[width=14cm]{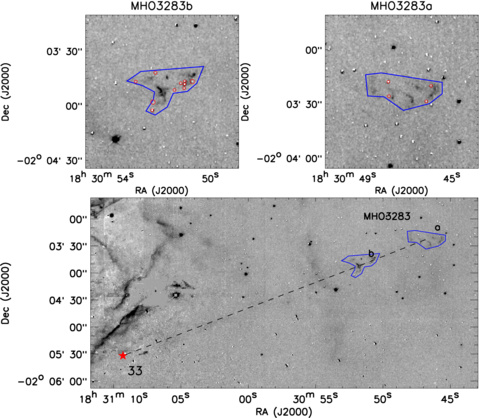}
\caption{Same as Fig.~\ref{figa02}, but for the region of MHO 3283. Note that the red circles mark the holes inside the polygons.}
\label{figa33}
\end{figure}

\begin{figure}
\includegraphics[width=14cm]{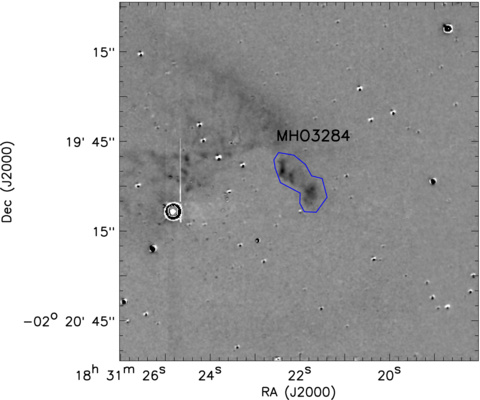}
\caption{Same as Fig.~\ref{figa01}, but for the region of MHO 3284. Note that we did not obtain the possible driving source of MHO 3284 due to the lack of YSOs in the ambient.}
\label{figa34}
\end{figure}

\begin{figure}
\includegraphics[width=14cm]{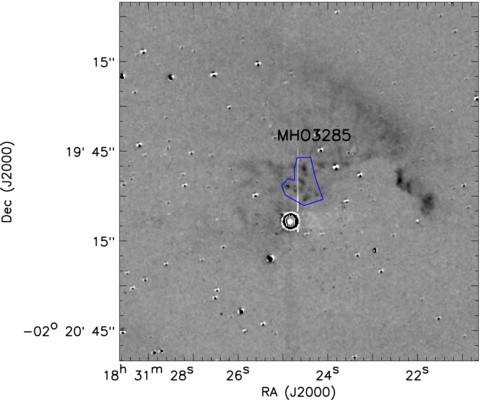}
\caption{Same as Fig.~\ref{figa01}, but for the region of MHO 3285. Note that we did not obtain the possible driving source of MHO 3285 due to the lack of YSOs in the ambient.}
\label{figa35}
\end{figure}

\begin{landscape}
\begin{figure}
\includegraphics[width=20cm]{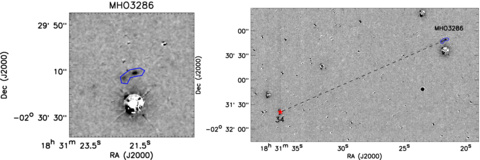}
\caption{Same as Fig.~\ref{figa02}, but for the region of MHO 3286.}
\label{figa36}
\end{figure}
\end{landscape}

\begin{landscape}
\begin{figure}
\includegraphics[width=20cm]{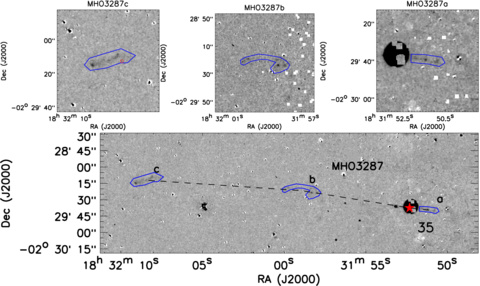}
\caption{Same as Fig.~\ref{figa02}, but for the region of MHO 3287. Note that the red circle marks the hole inside the polygon.  The circular bright blot in the corresponding position of ID\#35 is due to the ``persistence" instrument effect.}
\label{figa37}
\end{figure}
\end{landscape}

\begin{figure}
\includegraphics[width=14cm]{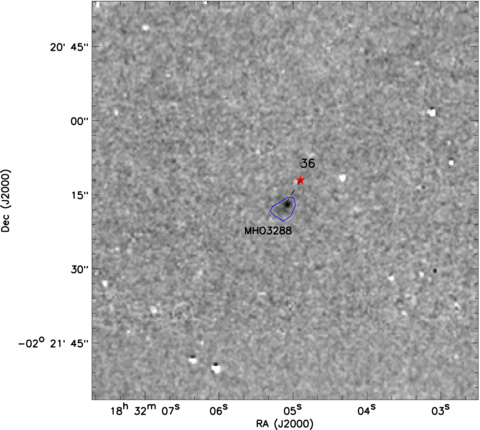}
\caption{Same as Fig.~\ref{figa01}, but for the region of MHO 3288.}
\label{figa38}
\end{figure}

\begin{figure}
\includegraphics[width=14cm]{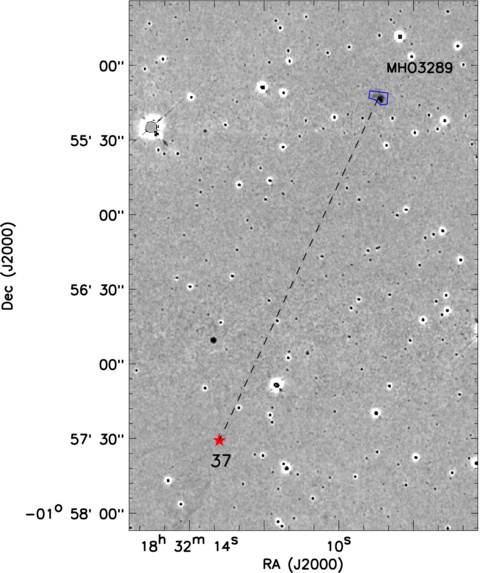}
\caption{Same as Fig.~\ref{figa01}, but for the region of MHO 3289.}
\label{figa39}
\end{figure}

\begin{figure}
\includegraphics[width=14cm]{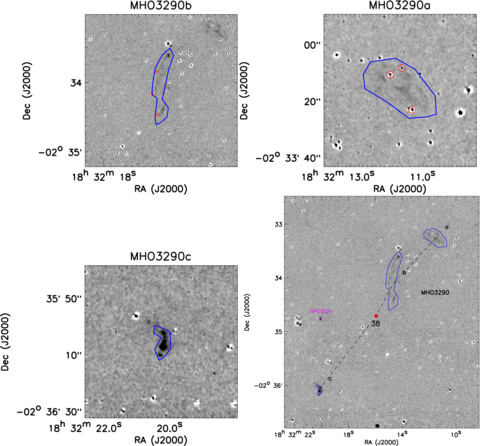}
\caption{Same as Fig.~\ref{figa01}, but for the region of MHO 3290. Note that the red circles mark the holes inside the polygons.}
\label{figa40}
\end{figure}

\begin{figure}
\includegraphics[width=14cm]{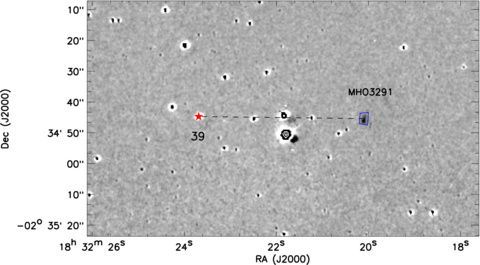}
\caption{Same as Fig.~\ref{figa01}, but for the region of MHO 3291.}
\label{figa41}
\end{figure}

\begin{figure}
\includegraphics[width=14cm]{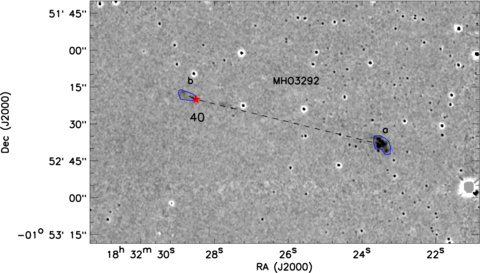}
\caption{Same as Fig.~\ref{figa01}, but for the region of MHO 3292.}
\label{figa42}
\end{figure}

Finally we have identified 45 MHOs which consist of 108 MHO features in the Aquila molecular cloud. Table~\ref{mhofeatures} shows the information of these MHO features, including their positions and areas. Note that we use a polygon to define the morphology of each MHO feature. The position and area of each MHO feature in table~\ref{mhofeatures} are calculated based on the polygon. We also present a continuum-subtracted $H_2$ image and give a brief description for each MHO. Figs.~\ref{figa01}-\ref{figa42} show these continuum-subtracted $H_2$ images. The blue and pink polygons mark the MHO features in the figures. Note that for some MHO features there are holes that are labeled with red circles in the polygons. The MHO features which belong to the same $H_2$ outflow and their possible driving source are connected with black dashed lines in the figures. 

\begin{itemize}
\item \myeph{MHO 2213}: Fig.~\ref{figa01} shows the region of MHO 2213 which consists of three features, MHO 2213a-c. \citet{mho2213} firstly detected this object, however, they only detected MHO 2213a due to the small sky coverage ($\sim$1\arcmin$\times$1\arcmin) of the observation. \citet{mho2213} also identified an infrared nebula which is associated with IRAS 18264-0143 and visible in our $K_s-$band image (see Fig.~\ref{nebshow}). We have identified a YSO, ID\#3, which is also associated with IRAS 18264-0143. It seems that MHO 2213 and ID\#3 constitute a bipolar outflow system. Therefore, we suggest ID\#3 as the possible driving source of MHO 2213.

\item \myeph{MHO 2214}: \citet{mho2213} firstly detected this object, but only the feature that is labeled as MHO 2214b in Fig.~\ref{figa02}. \citet{tei12} increased the number of MHO features up to two, MHO 2214b and MHO 2214r, corresponding to MHO 2214b and MHO 2214r5 in Fig.~\ref{figa02}, respectively. Using our deep $H_2$ images, we have identified six features in MHO 2214, MHO 2214r1-r5 and MHO 2214b. \citet{tei12} suggested a protostar, ``P0"\citep{bontemps10}, as the possible driving source of MHO 2214 because ``P0" and MHO 2214 are nearly located on a line. We identify a YSO, ID\#9, which is associated with ``P0". Thus we accept the conclusion of \citet{tei12} and suggest ID\#9 as the possible driving source of MHO 2214.

\item \myeph{MHO 3247}: Fig.~\ref{figa03} shows the region of MHO 3247 that consists of three features, MHO 3247a-c. \citet{tei12} firstly identified MHO 3247, but they only detected the northeast lobe, corresponding to MHO 3247c in Fig.~\ref{figa03}. Using our deep $H_2$ images, we detect the southwest lobe of MHO 3247 that is labeled as MHO 3247a and b. \citet{tei12} suggested the protostar, ``P2"\citep{bontemps10}, as the possible driving source of MHO 3247. We also identify a YSO, ID\#21, which is associated with ``P2". Thus we accept the conclusion of \citet{tei12} and also suggest ID\#21 as the possible driving source of MHO 3247.

\item \myeph{MHO 3248 and MHO 3251b2}: Fig.~\ref{figa04} shows the region of MHO 3248 and MHO 3251b2. \citet{tei12} firstly detected part of features of MHO 3248 which mainly corresponds to MHO 3248a in Fig.~\ref{figa04}. They suggested the Class 0 source, ``P3", as the driving source of MHO 3248. We detect another lobe of MHO 3248, labeled as MHO 3248b-c. We also identify a YSO, ID\#22, which is associated with ``P3". Thus we accept the suggestion of \citet{tei12} and suggest ID\#22 as the possible driving source of MHO 3248. MHO 3251b2 is firstly detected by \citet{tei12}, too. \citet{tei12} suggested that MHO 3251b2 may result from the interaction of two different outflows, among which one consists of ``P3" and MHO 3248 and the other consists of MHO 3251 (see the description about MHO 3251). MHO 3251b2 shows more detailed structures in our deep $H_2$ image. Based on the morphology of MHO 3251b2, we propose that MHO 3251b2 belongs to the outflow that is driven by ID\#22. Thus we suggest that MHO 3248, MHO 3251b2, and ID\#22 constitute a bipolar outflow system.

\item \myeph{MHO 3249}: \citet{tei12} firstly detected this object and suggested the Class I source, ``P4"\citep{bontemps10}, as the possible driving source. We also identify a YSO, ID\#24 that is associated with ``P4". Thus here we suggest ID\#24 as the driving source of MHO 3249, too. Note that we just mark the heads of two lobes of the outflow that is driven by ID\#24 in Fig.~\ref{figa05} with the blue polygons. In fact, \citet{tei12} suggested that there may be interaction between MHO 3248 and the blueshifted lobe of MHO 3249. They also suggested that part of MHO 3248a may belong to the outflow that consists of MHO 3249 and ID\#24. MHO 3248a shows more detailed structures in our deep $H_2$ image and obviously consists of two components in Fig.~\ref{figa04} and Fig.~\ref{figa05}: a north-south component may belong to the outflow driven by ID\#22; a NE-SW component should belong to the outflow driven by ID\#24. However, these two components are overlapped and hard to distinguish from each other. Thus we do not mark the sub-structures of MHO 3248a in Fig.~\ref{figa04}. Note especially that we assume that the whole MHO 3248a belongs to the outflow driven by ID\#22 when we calculate the outflow parameters such as jet length, jet opening angle, and jet position angle. 

\item \myeph{MHO 3250}: \citet{tei12} firstly detected this object, but only the feature that is marked as MHO 3250c in Fig.~\ref{figa06}. We identify more features, including MHO 3250a, b and d. \citet{tei12} suggested a young star, ``Y1" \citep{bontemps10}, as the driving source of MHO 3250. We have identified a YSO, ID\#18, which is associated with ``Y1". Thus here we accept the conclusion of \citet{tei12} and suggest that MHO 3250 is driven by ID\#18. 

\item \myeph{MHO 3251}: Fig.~\ref{figa07} shows the region of MHO 3251. \citet{tei12} firstly detected this object and suggested the Class I source, ``Y2" \citep{bontemps10} that is associated with our identified YSO ID\#17, as the possible driving source. \citet{tei12} suggested that the redshifted lobe of the outflow driven by ID\#17 could extend to MHO 3250c (see Fig.~\ref{figa06}) and there may be interaction between the outflow driven by ID\#17 and the outflow driven by ID\#18 (see the description about MHO 3250). MHO 3250c shows more detailed structures in our deep $H_2$ image (see Fig.~\ref{figa06}) and we argue that there is no obvious indication of interaction based on the morphology of MHO 3250c. Moreover, we detect a new feature, MHO 3251r, being located between ID\#17 and MHO 3250c. Thus we suggest MHO 3251r as the outskirt of the redshifted lobe of the outflow driven by ID\#17. On the other hand, \citet{tei12} suggested that the blueshifted lobe of the outflow driven by ID\#17 could extend to MHO 3251b2. However, we argue that MHO 3251b2 is likely to belong to the outflow driven by ID\#22 (see the description about MHO 3248). We detect a new feature, MHO 3251b1 that shows a highly collimated jet-like structure between ID\#17 and MHO 3251b2 in Fig.~\ref{figa07}. Therefore, we suggest MHO 3251b1 as the outskirt of the blueshifted lobe of the outflow driven by ID\#17. 

\item \myeph{MHO 3252}: \citet{tei12} firstly detected this object, but only the features that correspond to MHO 3252b1-b2, and r4 in Fig.~\ref{figa08}. We identify three new features in the west of MHO 3252r4, MHO 3252r1-r3. \citet{tei12} suggested the protostar, ``Y3" \citep{bontemps10}, as the possible driving source of MHO 3252. We detected an infrared source, ID\#19, as the counterpart of protostar ``Y3". ID\#19 is visible at 2-24\micron, but it is saturated in all IRAC and MIPS 24\,\micron~bands. Based on the good alignment of MHO 3252 and ID\#19, we suggest ID\#19 as the driving source of MHO 3252.

\item \myeph{MHO 3253}: Fig.~\ref{figa09} shows the region of MHO 3253. \citet{tei12} firstly detected this object, but only the feature corresponding to MHO 3253a in Fig.~\ref{figa09}. We detect a new feature, MHO 3253b, in the east of MHO 3253a. \citet{tei12} suggested the Class I source ``Y4" \citep{bontemps10}, which is associated with our identified YSO ID\#28, as the driving source of MHO 3253. Here we accept the suggestion of \citet{tei12} and propose ID\#28 as the driving source of MHO 3253.

\item \myeph{MHO 3254 and MHO 3277-3278}: Fig.~\ref{figa09} also shows the region of MHO 3254 and MHO 3277-3278. MHO 3254 was firstly identified by \citet{tei12}, but \citet{tei12} only detected two features that correspond to MHO 3254c in Fig.~\ref{figa09} and MHO 3254a in Fig.~\ref{figa10}. We detect a new feature, MHO 3254b, lying between MHO 3254a and c. \citet{tei12} suggested a YSO that is located to the east of ID\#28, ``Y5", as the possible driving source of MHO 3254a and c. We argue that MHO 3254c shows bow-like structure in our deep $H_2$ image and seems to be able to constitute, together with MHO 3253 and ID\#28, a bipolar outflow system. We also detect two new MHOs in the southwest of ID\#28, MHO 3277 and MHO 3278. It seems that MHO 3253, ID\#28, MHO 3254b-c, and MHO 3277-3278 align well with each other. Thus here we propose another possibility that MHO 3253, MHO 3254b-c, and MHO 3277-3278 are driven by ID\#28. Moreover, based on the location of MHO 3254a, we propose that MHO 3254a belongs to the outflow driven by ID\#23 (see the description about MHO 3255).

\item \myeph{MHO 3255}: \citet{tei12} firstly detected this object, but only the features corresponding to MHO 3255a and b in Fig.~\ref{figa10}. We identify a new feature, MHO 3255c, which is located $\sim$4\arcmin~to the northeast of MHO 3255b. \citet{tei12} suggested a protostar ``P5" that is associated with our identified YSO, ID\#23, as the possible driving source of MHO 3255. Based on the good alignment among MHO 3255, MHO 3254a, and ID\#23, we suggest that ID\#23 drives a bipolar outflow that consists of MHO 3255 and MHO 3254a.

\item \myeph{MHO 3259}: Fig.~\ref{figa11} shows the region of MHO 3259 that exhibits a bow-like structure. ID\#1 is the nearest YSO, lying in the southwest of MHO 3259. We propose ID\#1 as the possible driving source of MHO 3259.

\item \myeph{MHO 3260}: Fig.~\ref{figa12} shows the region of MHO 3260 which consists of four features. MHO 3260c and d are two knots. MHO 3260a and b show approximate point symmetric structures, which indicates that the driving source should be located between them. \citet{bontemps10} detected a protostar which is labeled as ID\#2 in Fig.~\ref{figa12}. Note that ID\#2 is invisible at 3.6-24\micron, but visible in Herschel PACS and SPIRE images, which indicates that ID\#2 may stay at a very early evolutionary stage. We suggest ID\#2 as the driving source of MHO 3260.

\item \myeph{MHO 3261}: Fig.~\ref{figa13} shows the region of MHO 3261 which consists of two features. The YSO ID\#4 is located at the northeast of MHO 3261b and associated with an infrared nebula (see Fig.~\ref{nebshow}). The infrared nebula exhibits the NE-SW elongated structure. We suggest ID\#4 as the driving source of MHO 3261.

\item \myeph{MHO 3262}: This object is an elongated knot in Fig.~\ref{figa14}. ID\#5 is the nearest YSO to it and we suggest ID\#5 as the possible driving source of MHO 3262.

\item \myeph{MHO 3263}: Fig.~\ref{figa15} shows the region of MHO 3263 which consists of three features. The YSO ID\#6 and MHO 3263 are roughly located on a straight line. Thus we suggest ID\#6 as the possible driving source of MHO 3263.

\item \myeph{MHO 3264}: Fig.~\ref{figa16} shows the region of MHO 3264 which consists of five features. The YSO ID\#7 is roughly aligned with the features of MHO 3264. ID\#7 is also associated with an infrared nebula (see Fig.~\ref{nebshow}) which seems to trace the cavity wall of the west lobe of outflow. Thus we suggest ID\#7 as the possible driving source of MHO 3264.

\item \myeph{MHO 3265}: Fig.~\ref{figa17} shows the region of MHO 3265 which consists of two features. The YSO ID\#8 is located between MHO 3265a and b. Note that we suggest MHO 3265a belongs to another outflow rather than the outflow that consists of MHO 2214 based on its morphology although MHO 3265a is located between MHO 2214r3 and MHO 2214r4 (see the description about MHO 2214). Considering the good alignment of ID\#8 and MHO 3265, we propose ID\#8 as the possible driving source of MHO 3265.

\item \myeph{MHO 3266}: MHO 3266 shows a complex structure in Fig.~\ref{figa18}. The YSO ID\#10 is located in the southwest of MHO 3266 and seems to be located on the extension line of the major axis of MHO 3266. Thus we suggest ID\#10 as the possible driving source of MHO 3266.

\item \myeph{MHO 3267}: Fig.~\ref{figa19} shows the region of MHO 3267 which consists of three features. The YSO ID\#11 is associated with IRAS 18270-0153 \citep{mho2213,connelley10} and an infrared nebula \citep{mho2213}. This infrared nebula shows a NE-SW bipolar structure in our $K_s$ image (see Fig.~\ref{nebshow}). It seems that ID\#11, the infrared nebula, and MHO 3267 constitute a bipolar outflow system. 

\item \myeph{MHO 3268}: This object consists of four features, MHO 3268a-d, in an arrangement of a fan shape with the center of ID\#12 in Fig.~\ref{figa20}. ID\#12 is also associated with a protostar identified by \citet{bontemps10} using the Herschel data. Here we suggest ID\#12 as the possible driving source of MHO 3268.

\item \myeph{MHO 3269}: Fig.~\ref{figa21} shows the region of MHO 3269. The YSO ID\#13 is located to the east of MHO 3269 and associated with an infrared nebula that seems to trace the cavity walls of an east-west outflow (see Fig.~\ref{nebshow}). MHO 3269 and ID\#13 are roughly on a straight line. Thus we suggest ID\#13 as the possible driving source of MHO 3269.

\item \myeph{MHO 3270}: Fig.~\ref{figa22} shows the region of MHO 3270 that consists of two features. MHO 3270a seems to consist of a knot and a jet-like structure driven by the YSO ID\#14 while MHO 3270b shows a fan-shaped structure. ID\#14 and MHO 3270 align well with each other. Thus we suggest ID\#14 as the possible driving source of MHO 3270.

\item \myeph{MHO 3271}: Fig.~\ref{figa23} shows the region of MHO 3271. There are several YSOs in the ambient of MHO 3271. Among them, ID\#15 is associated with an infrared nebula (see Fig.~\ref{nebshow}). It seems that ID\#15, the infrared nebula, and MHO 3271 constitute a NE-SW bipolar outflow system and the infrared nebula represents the northeast lobe while MHO 3271 traces the southwest lobe.

\item \myeph{MHO 3272}: MHO 3272 shows a complex structure in Fig.~\ref{figa24}. ID\#16 is the nearest YSO. It seems that ID\#16 and MHO 3272 constitute a bipolar outflow system.

\item \myeph{MHO 3273}: Fig.~\ref{figa25} shows the region of MHO 3273 which consists of eight features. The YSO ID\#20 is located in the vicinity of MHO 3273d. ID\#20 and MHO 3273 are located roughly on a straight line. Based on the morphologies and locations, we suggest ID\#20 as the possible driving source of MHO 3273.

\item \myeph{MHO 3274}: Fig.~\ref{figa26} shows the region of MHO 3274. ID\#25 is the nearest YSO and associated with an infrared nebula which is shown in Fig.~\ref{nebshow}. It seems that ID\#25, the infrared nebula, and MHO 3274 constitute a bipolar outflow system. Thus we suggest ID\#25 as the driving source of MHO 3274.

\item \myeph{MHO 3275}: This object is an elongated knot in Fig.~\ref{figa27}. ID\#25 (see Fig.~\ref{figa26}) is the nearest YSO, but ID\#25 drives an east-west outflow (see the description about MHO 3274), which disagrees with the location of MHO 3275. Therefore, we suggest the secondary nearest YSO, ID\#26, as the possible driving source of MHO 3275.

\item \myeph{MHO 3276}: This object exhibits the bow-like structure in Fig.~\ref{figa28}, which indicates that its possible driving source may be located to the southwest of it. ID\#27 is located in the southwest of MHO 3276 and also associated with a protostar identified by \citet{bontemps10} using the Herschel data. Thus we suggest ID\#27 as the possible driving source of MHO 3276.

\item \myeph{MHO 3279}: Fig.~\ref{figa29} shows the region of MHO 3279 that consists of two features. The YSO ID\#29 is located between the two features of MHO 3279. It seems that MHO 3279 and ID\#29 constitute a bipolar outflow system. Note that ID\#29 is also associated with a SE-NW infrared nebula (see Fig.~\ref{nebshow}). Thus we suggest ID\#29 as the driving source of MHO 3279.

\item \myeph{MHO 3280}: This object shows bow shock morphology in Fig.~\ref{figa30}, which indicates that its possible driving source may be located to its northwest. The YSO ID\#30 is located in the northwest of MHO 3280 and associated with an infrared nebula that exhibits SE-NW bipolar structure in Fig.~\ref{nebshow}. We suggest ID\#30 as the driving source of MHO 3280.

\item \myeph{MHO 3281}: Fig.~\ref{figa31} shows the region of MHO 3281 that consists of three features. The YSO ID\#31 is located between MHO 3281a and b. Note that ID\#31 is also associated with an infrared nebula (see Fig.~\ref{nebshow}). It seems that MHO 3281 and ID\#31 constitute a bipolar outflow system.

\item \myeph{MHO 3282}: This object seems to be a bright knot with a faint ring-like structure in Fig.~\ref{figa32}. ID\#32 is the nearest YSO and we suggest it as the possible driving source of MHO 3282.

\item \myeph{MHO 3283}: Fig.~\ref{figa33} shows the region of MHO 3283 that consists of two features. MHO 3283a exhibits the multiple bow shock morphology, which indicates that its possible driving source should be located to its east. The YSO ID\#33 in the southeast of MHO 3283 is located roughly on the line connecting MHO 3283a and MHO 3283b. Thus we suggest ID\#33 as the possible driving source of MHO 3283.

\item \myeph{MHO 3284 and MHO 3285}: MHO 3284 shows complex morphology in Fig.~\ref{figa34} while MHO 3285 consists of some knots in Fig.~\ref{figa35}. No YSOs have been identified in the ambient of MHO 3284 and MHO 3285. Therefore, we can not infer any driving source candidates for MHO 3284 and MHO 3285.

\item \myeph{MHO 3286}: This object is a chain of two bright knots in Fig.~\ref{figa36}, which indicates that its possible driving source may be located to its southeast or northwest. There is a YSO, ID\#34, to the southeast of MHO 3286. ID\#34 and the knots of MHO 3286 seems to align with each other. Thus we suggest ID\#34 as the possible driving source of MHO 3286.

\item \myeph{MHO 3287}: Figure~\ref{figa37} shows the region of MHO 3287 that consists of three features. MHO 3287a is a collimated jet in the vicinity of the YSO ID\#35 that is associated with a protostar identified by \citet{bontemps10} using the Herschel data. There is good alignment between ID\#35 and MHO 3287. Thus we suggest ID\#35 as the driving source of MHO 3287. Note that the circular bright blot in the corresponding position of ID\#35 is due to the instrument effect, ``persistence", which occurs due to charges being trapped in high flux or saturated regions in one exposure that are then slowly released in subsequent exposures. Since exposures are dithered, the released charges appear as new ``artificial sources" in the subsequent images.

\item \myeph{MHO 3288}: This object is a patch in Fig.~\ref{figa38}. The nearest YSO to MHO 3288 is ID\#36, which is also associated with an infrared nebula (see Fig.~\ref{nebshow}). We suggest ID\#36 as the possible driving source of MHO 3288.

\item \myeph{MHO 3289}: This object shows a bow-like structure in Fig.~\ref{figa39}, which indicates that the driving source of MHO 3289 may be located to the southeast of it. ID\#37 is located to the southeast of MHO 3289 and associated with a SE-NW bipolar infrared nebula (see Fig.~\ref{nebshow}). We suggest ID\#37 as the possible driving source of MHO 3289.

\item \myeph{MHO 3290}: This object consists of three features that align well with each other in Fig.~\ref{figa40}. The YSO ID\#38 is located between MHO 3290b and MHO 3290c and associated with an infrared nebula that extends towards southeast from ID\#38 in Fig.~\ref{nebshow}. It seems that ID\#38 and MHO 3290 constitute a bipolar outflow system. We suggest ID\#38 as the driving source of MHO 3290.

\item \myeph{MHO 3291}: This object is a knot in Fig.~\ref{figa41}. ID\#39 is the nearest YSO to MHO 3291. We suggest ID\#39 as the possible driving source of MHO 3291.

\item \myeph{MHO 3292}: Figure~\ref{figa42} shows the region of MHO 3292. The YSO ID\#40 is located roughly on the line connecting the features of MHO 3292. ID\#40 is also associated with an infrared nebula that extends towards the northeast from ID\#40 in Fig.~\ref{nebshow}. Thus we suggest ID\#40 as the possible driving source of MHO 3292.

\end{itemize}
\begin{deluxetable}{lccccc}
\centering
\tabletypesize{\scriptsize}
\tablewidth{0pt}
\tablecaption{MHO features in Aquila.\label{mhofeatures}}
\tablehead{
\colhead{}&
\colhead{RA\tablenotemark{a}}&
\colhead{Dec\tablenotemark{a}}&
\colhead{Area\tablenotemark{b}}&
\colhead{Radius\tablenotemark{c}}&
\colhead{$H_2~1-0~S(1)$ flux}\\
\colhead{MHO}&
\colhead{(J2000)}&
\colhead{(J2000)}&
\colhead{(arcsec$^2$)}&
\colhead{(arcsec)}&
\colhead{(10$^{-18}$ W m$^{-2}$)}
}
\startdata
3259& 18 28 41.4  &-01 32 08&24.1&2.8&19.31$\pm$1.85\\
3261a& 18 28 52.8  &-01 46 48&9.5&1.7&2.11$\pm$0.68\\
3260a& 18 29 03.0  &-01 39 10&31.2&3.2&10.57$\pm$0.89\\
3260b& 18 29 04.4  &-01 39 01&33.6&3.3&27.72$\pm$1.98\\
2213a& 18 29 04.9  &-01 41 52&61.9&4.4&33.70$\pm$2.61\\
3260c& 18 29 05.4  &-01 38 52&5.8&1.4&0.69$\pm$0.13\\
3260d& 18 29 06.2  &-01 38 47&5.4&1.3&1.00$\pm$0.16\\
2213b& 18 29 07.9  &-01 42 16&26.5&2.9&6.49$\pm$0.64\\
2213c& 18 29 08.8  &-01 42 34&57.8&4.3&29.99$\pm$2.35\\
3261b& 18 29 08.9  &-01 46 27&34.5&3.3&4.83$\pm$1.48\\
3262& 18 29 14.1  &-01 57 59&13.3&2.1&1.87$\pm$0.62\\
3252r1& 18 29 15.8  &-02 13 21&44.3&3.8&10.08$\pm$1.75\\
3264a& 18 29 17.8  &-01 39 17&50.0&4.0&16.65$\pm$3.96\\
3268a& 18 29 18.9  &-02 19 37&370.1&10.9&11.30$\pm$2.34\\
3264b& 18 29 20.6  &-01 38 59&102.7&5.7&10.63$\pm$2.67\\
3263a& 18 29 21.2  &-01 37 33&9.1&1.7&1.23$\pm$0.37\\
3263b& 18 29 22.5  &-01 37 52&32.5&3.2&3.71$\pm$1.01\\
3268b& 18 29 24.9  &-02 19 36&571.3&13.5&75.38$\pm$11.76\\
2214r1& 18 29 25.6  &-01 49 55&40.8&3.6&2.18$\pm$0.41\\
3264c& 18 29 25.7  &-01 38 59&28.8&3.0&2.75$\pm$0.80\\
3268c& 18 29 26.5  &-02 18 28&254.8&9.0&219.70$\pm$32.37\\
3263c& 18 29 26.9  &-01 39 49&17.6&2.4&3.35$\pm$0.90\\
2214r2& 18 29 27.8  &-01 50 03&61.5&4.4&5.26$\pm$0.78\\
3264d& 18 29 29.3  &-01 38 41&39.3&3.5&6.80$\pm$1.72\\
3252r2& 18 29 30.1  &-02 12 20&121.1&6.2&59.08$\pm$9.05\\
3268d& 18 29 30.5  &-02 16 27&59.0&4.3&8.69$\pm$1.64\\
3273a& 18 29 32.5  &-01 32 35&251.1&8.9&64.67$\pm$14.59\\
3252r3& 18 29 33.0  &-02 11 52&103.1&5.7&20.26$\pm$3.34\\
2214r3& 18 29 33.3  &-01 50 33&81.2&5.1&12.67$\pm$1.59\\
3269a& 18 29 33.6  &-01 33 12&34.5&3.3&5.91$\pm$1.52\\
3265a& 18 29 34.0  &-01 50 28&25.5&2.8&8.35$\pm$1.08\\
2214r4& 18 29 34.8  &-01 50 31&43.8&3.7&15.19$\pm$1.83\\
3267a& 18 29 36.4  &-01 52 00&5.0&1.3&0.67$\pm$0.14\\
2214r5& 18 29 36.4  &-01 50 43&129.9&6.4&38.15$\pm$4.27\\
3267b& 18 29 36.9  &-01 51 38&8.4&1.6&2.48$\pm$0.38\\
3269b& 18 29 38.2  &-01 33 02&36.7&3.4&16.19$\pm$3.81\\
2214b& 18 29 39.7  &-01 51 27&186.1&7.7&108.30$\pm$11.38\\
3265b& 18 29 39.7  &-01 52 46&312.6&10.0&10.11$\pm$1.53\\
3267c& 18 29 40.1  &-01 51 08&46.0&3.8&5.75$\pm$0.82\\
3269c& 18 29 40.3  &-01 32 58&19.4&2.5&4.75$\pm$1.23\\
3250a& 18 29 43.2  &-02 05 29&4.9&1.2&1.74$\pm$0.36\\
3264e& 18 29 45.0  &-01 37 36&18.0&2.4&9.69$\pm$2.35\\
3270b& 18 29 47.0  &-01 59 34&57.8&4.3&18.95$\pm$1.20\\
3270a& 18 29 47.0  &-01 55 52&8.9&1.7&2.73$\pm$0.26\\
3250b& 18 29 50.8  &-02 05 08&58.1&4.3&4.75$\pm$0.93\\
3273b& 18 29 50.9  &-01 34 38&294.3&9.7&43.96$\pm$10.16\\
3266& 18 29 53.0  &-01 29 27&436.9&11.8&204.70$\pm$55.70\\
3273c& 18 29 55.3  &-01 35 38&102.2&5.7&17.63$\pm$4.18\\
3250c& 18 29 56.8  &-02 04 33&333.1&10.3&95.79$\pm$14.45\\
3271& 18 29 57.3  &-02 02 28&44.4&3.8&3.30$\pm$0.66\\
3278& 18 29 57.8  &-02 12 55&32.0&3.2&7.62$\pm$1.35\\
3252r4& 18 29 58.2  &-02 10 52&243.0&8.8&70.21$\pm$10.72\\
3272a& 18 29 59.2  &-02 25 53&216.5&8.3&74.74$\pm$10.09\\
3247a& 18 30 00.0  &-02 03 12&23.0&2.7&9.79$\pm$1.66\\
3251r& 18 30 00.1  &-02 05 26&102.6&5.7&11.29$\pm$1.98\\
3272b& 18 30 00.7  &-02 24 08&137.8&6.6&44.77$\pm$6.16\\
3249b& 18 30 00.8  &-02 04 59&83.5&5.2&73.10$\pm$11.07\\
3252b1& 18 30 01.8  &-02 10 24&29.0&3.0&50.41$\pm$7.73\\
3251b1& 18 30 02.0  &-02 06 28&7.5&1.5&1.06$\pm$0.26\\
3273d& 18 30 02.8  &-01 36 30&30.7&3.1&8.16$\pm$2.03\\
3247b& 18 30 03.0  &-02 02 59&123.6&6.3&192.30$\pm$28.32\\
3277& 18 30 03.1  &-02 12 18&94.5&5.5&9.72$\pm$1.74\\
3248a& 18 30 03.5  &-02 03 57&1782.2&23.8&807.50$\pm$116.70\\
3248b& 18 30 03.8  &-02 01 48&114.4&6.0&20.75$\pm$3.36\\
3255a& 18 30 03.9  &-02 11 11&50.4&4.0&16.18$\pm$2.68\\
3247c& 18 30 04.2  &-02 02 26&225.2&8.5&155.50$\pm$23.05\\
3248c& 18 30 04.2  &-02 02 05&164.3&7.2&24.79$\pm$3.99\\
3255b& 18 30 05.7  &-02 10 52&308.7&9.9&308.80$\pm$45.18\\
3251b2& 18 30 05.8  &-02 08 03&1799.0&23.9&372.00$\pm$54.64\\
3249r& 18 30 07.8  &-02 01 04&64.8&4.5&12.56$\pm$2.14\\
3254a& 18 30 08.3  &-02 10 49&62.3&4.5&21.92$\pm$3.54\\
3273e& 18 30 08.7  &-01 37 44&18.8&2.4&2.44$\pm$0.71\\
3273f& 18 30 11.1  &-01 38 09&17.6&2.4&1.34$\pm$0.44\\
3252b2& 18 30 14.1  &-02 09 22&2816.7&29.9&316.00$\pm$46.85\\
3250d& 18 30 16.1  &-02 01 52&294.1&9.7&25.90$\pm$4.21\\
3274& 18 30 16.4  &-02 07 27&28.2&3.0&1.98$\pm$0.46\\
3275& 18 30 16.7  &-02 06 28&10.4&1.8&1.40$\pm$0.31\\
3254b& 18 30 17.8  &-02 10 58&128.1&6.4&20.26$\pm$3.37\\
3255c& 18 30 21.8  &-02 09 56&68.8&4.7&5.36$\pm$1.07\\
3254c& 18 30 22.6  &-02 10 57&108.8&5.9&33.86$\pm$5.33\\
3253a& 18 30 27.2  &-02 10 35&26.6&2.9&18.44$\pm$2.98\\
3253b& 18 30 28.1  &-02 10 26&33.3&3.3&2.45$\pm$0.54\\
3279a& 18 30 28.3  &-01 55 59&10.8&1.9&3.37$\pm$0.95\\
3276& 18 30 28.9  &-01 48 19&37.5&3.5&13.63$\pm$3.21\\
3279b& 18 30 29.1  &-01 56 07&22.8&2.7&2.11$\pm$0.65\\
3273g& 18 30 43.9  &-01 42 05&106.8&5.8&14.81$\pm$3.60\\
3273h& 18 30 46.5  &-01 43 29&28.2&3.0&7.61$\pm$1.90\\
3283a& 18 30 46.9  &-02 03 23&578.8&13.6&55.04$\pm$10.94\\
3281a& 18 30 48.2  &-01 56 32&16.9&2.3&6.35$\pm$0.84\\
3281b& 18 30 49.0  &-01 55 45&15.6&2.2&3.15$\pm$0.48\\
3281c& 18 30 50.4  &-01 55 38&41.4&3.6&5.61$\pm$0.79\\
3283b& 18 30 51.8  &-02 03 52&487.6&12.5&75.01$\pm$14.60\\
3280& 18 30 55.1  &-01 57 37&32.8&3.2&14.09$\pm$1.68\\
3282& 18 30 59.4  &-02 29 19&72.6&4.8&16.27$\pm$2.13\\
3286& 18 31 21.7  &-02 30 12&35.6&3.4&10.59$\pm$1.72\\
3284& 18 31 22.0  &-02 19 58&186.8&7.7&96.04$\pm$10.74\\
3285& 18 31 24.6  &-02 19 55&131.8&6.5&58.25$\pm$6.68\\
3287a& 18 31 51.1  &-02 29 40&80.6&5.1&6.71$\pm$1.11\\
3287b& 18 31 58.9  &-02 29 22&200.1&8.0&10.99$\pm$1.74\\
3288& 18 32 05.1  &-02 21 18&15.5&2.2&1.61$\pm$0.26\\
3287c& 18 32 08.5  &-02 29 12&211.0&8.2&17.84$\pm$2.55\\
3289& 18 32 08.9  &-01 55 13&29.9&3.1&10.18$\pm$0.88\\
3290a& 18 32 11.5  &-02 33 16&341.6&10.4&25.68$\pm$2.91\\
3290b& 18 32 14.7  &-02 34 07&595.7&13.8&38.09$\pm$4.16\\
3291& 18 32 20.1  &-02 34 45&10.0&1.8&3.22$\pm$0.44\\
3290c& 18 32 20.2  &-02 36 05&47.5&3.9&16.49$\pm$1.82\\
3292a& 18 32 23.4  &-01 52 39&35.8&3.4&19.24$\pm$1.87\\
3292b& 18 32 28.7  &-01 52 19&20.7&2.6&3.21$\pm$0.43\\
\enddata
\tablenotetext{a}{Center of the polygon that is identifited based on the morphology and surface brightness distribution of each MHO feature.}
\tablenotetext{b}{Area of the polygon.}
\tablenotetext{c}{Equivalent circular radius of the polygon.}
\end{deluxetable}

\clearpage
\clearpage
\clearpage
\clearpage
\clearpage
\clearpage
\clearpage
\clearpage
\clearpage
\clearpage
\clearpage
\clearpage
\clearpage
\clearpage
\clearpage
\clearpage
\clearpage
\clearpage
\clearpage
\clearpage
\clearpage
\clearpage
\clearpage
\clearpage
\clearpage
\clearpage
\clearpage
\clearpage
\clearpage
\clearpage
\clearpage
\clearpage
\clearpage
\clearpage
\clearpage
\clearpage
\clearpage
\clearpage
\clearpage
\clearpage
\clearpage
\clearpage

\clearpage

\end{document}